\documentclass[aps,pra, groupedaddress,nofootinbib,showpacs,twocolumn]{revtex4-1}

\usepackage{graphicx}
\usepackage[labelfont=small]{subcaption}
\usepackage{amsmath}

\graphicspath{{figures/}}

\newcommand{\dist}[1]{f_{#1}(#1)}

\begin{document}

\title{Energy distributions of an ion in a radiofrequency trap immersed in a buffer gas under the influence of additional external forces}
 
\author{I. Rouse}
\affiliation{Department of Chemistry, University of Basel, Basel, Switzerland}
\author{S. Willitsch}

\affiliation{Department of Chemistry, University of Basel, Basel, Switzerland}

\date{\today}

\begin{abstract}
An ion held in a radiofrequency trap interacting with a uniform buffer gas of neutral atoms develops a steady-state energy distribution characterised by a power-law tail at high energies instead of the exponential decay characteristic of thermal equilibrium. We have previously shown that the Tsallis statistics frequently used as an empirical model for this distribution is a good approximation when the ion is heated due to a combination of micromotion interruption and exchange of kinetic energy with the buffer gas [I. Rouse and S. Willitsch, Phys. Rev. Lett. \textbf{118}, 143401 (2017)]. Here, we extend our treatment to include the heating due to additional motion of the ion caused by external forces, including the ``excess micromotion'' induced by uniform electric fields and rf phase offsets. We show that this also leads to a Tsallis distribution with a potentially different power-law exponent from that observed in the absence of this additional forced motion, with the difference increasing as the ratio of the mass of the neutral atoms to that of the ion decreases. Our results indicate that unless the excess micromotion is minimised to a very high degree, then  even a system with very light neutrals and a heavy ion does not exhibit a thermal distribution.  
\end{abstract}

\pacs{ }

\maketitle


\section{Introduction}
Ultracold atoms, ions and molecules are of great interest in atomic, molecular, and chemical physics, from testing fundamental concepts with precision spectrosopic measurements to investigating the nature of elementary chemical reactions \cite{carr09a,bell09b,  haerter14a, sias14a, willitsch15a,tomza17a,willitsch17a}. By reducing the kinetic energy of the particles to reach temperatures below 1~mK, cross sections and reaction rates can be measured with high resolution. For neutral atoms, the combination of Doppler, Sisyphus and evaporative cooling enables reaching temperatures as low as the nanokelvin regime. Charged particles, however, are usually trapped in much smaller numbers, and so evaporative cooling no longer offers a route to extremely low temperatures. It may seem  possible to prepare a sample of ultracold atoms at the desired temperature, and bring the ions into thermal contact with these atoms in a hybrid trapping setup to reduce their energy to an equally low temperature via cooling collisions \cite{haerter14a, sias14a, willitsch15a,tomza17a}. While it is true that in each collision the ion may transfer energy to the atom, there is a complication due to the experimental techniques usually employed to trap ions. If the charged particles are held in a radiofrequency (rf) trap, then the ion's motion consists of a spectrum of frequency components, containing low-frequency secular motion and high-frequency micromotion. A collision with a neutral atom leads to a randomisation of the phase and amplitude of this motion, and the outgoing trajectory of the ion may correspond to a higher average energy than the trajectory before the collision even if the ion's velocity is reduced to zero by the collision \cite{devoe09a,zipkes11a,chen14a}. 

This effect is typically referred to as micromotion interruption and has two important consequences. Firstly, the mean energy of the ion may be several times larger than that predicted if it was in thermal equilibrium with the atomic cloud, and secondly the ion's energy no longer follows a thermal distribution \cite{devoe09a,chen14a}. The observed distribution has frequently been empirically modelled using Tsallis statistics \cite{tsallis88a,devoe09a,meir16a},
\begin{equation} \label{eq:tsallisDist}
\dist{E} = \left(\frac{n_T}{\langle\beta\rangle}\right)^{-k-1}  \frac{\Gamma (k+n_T+1)}{\Gamma (k+1) \Gamma (n_T)}     \frac{E^k}{ \left(\frac{\langle\beta\rangle E}{n_T}+1\right)^{k + n_T + 1}}.
\end{equation}
where $E$ is the energy of the ion, $E^k$ represents the density of states ($k =2$ for a 3D harmonic oscillator), $\Gamma(x)$ is the Gamma function, $\langle \beta \rangle$ is a scale parameter, and the Tsallis exponent $n_T$ parameterises the degree of departure from thermal equilibrium, with $n_T \rightarrow \infty$ corresponding to thermal equilibrium. At high energy, this distribution exhibits an asymptotic decay of the form $E^{-(n_T +1)}$, while in the limit $n_T \rightarrow \infty$ Eq.~\eqref{eq:tsallisDist} converges to a thermal distribution. Tsallis statistics emerge as a limiting case in the formalism of superstatistics, in which a system is viewed as being in an instantaneous thermal equilibrium but with a fluctuating temperature, in which case $\langle \beta \rangle$ represents the mean value of the ``inverse temperature'' $\beta = 1/(k_B T)$ \cite{beck03a}. Regardless of the exact nature of the fluctuations of the temperature, it can be shown that Tsallis statistics arise as a first-order approximation to the energy distribution, and in the special case in which the temperature follows an inverse-Gamma distribution, this approximation becomes exact for all energies \cite{beck03a}. We have previously demonstrated that the inverse-Gamma distribution is a good approximation for the distribution of the secular temperature of an ion immersed in a uniform neutral buffer gas undergoing Langevin collisions, thus leading to Tsallis statistics for the ion's secular energy distribution \cite{rouse17a}. We now extend our treatment to include the case where there is an additional component of motion due to the presence of external forces, such as the ``excess micromotion'' (EMM) of the ion \cite{berkeland98a}. Such forces usually arise from experimental imperfections, e.g., electric fields from patch potentials, which are ubiquitous in realistic settings unless carefully compensated.  We present an overview of the motion of an ion in a rf quadrupole trap subject to an additional spatially-independent force and clarify the differences between the intrinsic micromotion due to the rf field and the additional motion induced by the combination of these external forces and the trapping potential, including in-phase excess micromotion as a special case. From this, we derive analytical expressions for the mean steady-state energy when the ion interacts with a neutral buffer gas which may be used to evaluate the effects of a wide range of forms of forced motion of the ion. Furthermore, we show that the presence of this motion alters the power-law exponent by contributing an additional source of additive noise, with the effect becoming more pronounced at low neutral-to-ion mass ratios. This has implications for, e.g., experiments employing lithium as a buffer gas \cite{joger17a,haze17a}, since unless excess micromotion is compensated to a very high degree it cannot be assumed that the ion will exhibit a thermal distribution. The results of our analytical model for the mean energies and values of $n_T$ are compared to numerical simulations, finding excellent agreement.

\section{Theory}
\subsection{Ion motion}
\subsubsection{Basic equations}
The classical motion of a single ion in a radiofrequency quadrupole trap with no other external forces has been described in detail elsewhere \cite{major05a, march05a, olver2010a} and so we provide only a brief description of the standard equations to establish notation. A list of the most commonly used symbols is given in Table~\ref{tab:notation}  for reference. The combination of two quadrupole potentials, one static and one proportional to $\cos (\Omega t)$, results in the homogenous Mathieu differential equation describing the motion of an ion in the trap,
\begin{equation} \label{eq:mathieuHomogenous}
\ddot r_j(\tau)  + [a_j - 2q_j \cos(2 \tau) ] r_j(\tau)= 0,
\end{equation}
where $\tau = \Omega t/2, j \in (x,y,z) $ and $a_j,q_j$ parameterise the static and time-varying fields respectively. We assume throughout that the values of $a_j$ and $q_j$ result in stable motion \cite{olver2010a}.  For $|q_j| \lesssim 0.1$, an approximate solution to Eq.~\eqref{eq:mathieuHomogenous} may be found by treating the ion as undergoing harmonic motion in a static pseudopotential superimposed by small high-frequency oscillations, referred to as the secular motion and micromotion, respectively \cite{major05a}. This adiabatic approximation is not accurate enough for our purposes and so we use the exact solutions to the Mathieu equation. We take the pair of fundamental solutions $\mathrm{ce}(a_j,q_j,\tau)$ and $\mathrm{se}(a_j,q_j,\tau)$ as defined in Ref.~\cite{olver2010a}, and denote these $\mathrm{ce_j}(\tau)$ and $\mathrm{se_j}(\tau)$ respectively. The solution to Eq.~\eqref{eq:mathieuHomogenous} is,
\begin{equation} \label{eq:mathieuPositionHomogenous}
r_j(\tau) = r_{h,j}(A_j,\phi_j,\tau) = A_j [ \mathrm{ce_j}(\tau) \cos \phi_j - \mathrm{se_j} (\tau)  \sin \phi_j],
\end{equation}
where we have parameterised the two constants of integration in terms of an amplitude $A_j$ and a phase angle $\phi_j$ by analogy to the  harmonic oscillator, and the index $h$ indicates that this is the solution to the homogenous equation. The function  $\mathrm{ce_j}(\tau)$  may be expanded into a  Fourier series of the form \cite{olver2010a},
\begin{equation} \label{eq:mathieuCE}
\mathrm{ce_j}(\tau) = \sum_m c_{2m,j} \cos[ (\beta_j + 2m)\tau],
\end{equation}
 where $\beta_j$ is the characteristic exponent. The coefficients are functions of $a_j,q_j$ and $m$, and are normalised such that $\sum_m c_{2m}^2 = 1$ \cite{olver2010a}. The series for $\mathrm{se_j}(\tau)$ is defined analogously to Eq.~\eqref{eq:mathieuCE} in terms of sine functions. Substituting these expressions into Eq.~\eqref{eq:mathieuPositionHomogenous} and simplfiying the result produces,
\begin{equation} \label{eq:mathieuPositionHomogenousCombined}
r_{h,j}(A_j,\phi_j,\tau) = A_j \sum_m c_{2m,j} \cos[ (\beta_j + 2m) \tau + \phi_j].
\end{equation}
The $m=0$ term of this series describes oscillations at the frequency of the secular motion of the adiabatic approximation \cite{major05a}. We therefore identify this term as the secular motion, which is   harmonic oscillations at the secular frequency $\omega_j = \frac{1}{2} \beta_j \Omega$ and amplitude $c_{0,j} A_j$. The remaining terms with $m \neq 0$ are motion at frequencies given by $\omega_j + m \Omega$ with amplitudes  $c_{2m} A_j$. Under typical trapping conditions , i.e., $q_j < 0.5$, the amplitude of these terms  is much smaller than the amplitude of the secular motion. Thus, these are collectively referred to as the  micromotion  of the ion \cite{major05a}. To distinguish this from the excess micromotion discussed in the next section, we adopt the convention that the micromotion proportional to $A_j$ is the intrinsic micromotion (IMM), and the sum of these terms and the secular motion is the intrinsic motion. As a result of the time-dependent trapping potential, the ion's energy is not a constant. However, we may define a time-conserved energy from the secular motion of the ion, i.e., the secular energy,
\begin{equation} \label{eq:secularEnergy}
E_j =  \frac{m_i}{2} \frac{\Omega^2}{4} A_j^2 \beta^2_j c_{0,j}^2 ,
\end{equation} 
where $m_i$ is the mass of the ion \cite{major05a}.

\begin{table}[]
\centering
\caption{Table of symbols}
\label{tab:notation}
\begin{tabular}{ll}
\hline
\hline
Symbol & Definition \\
\hline
$a_j, q_j $&  Mathieu stability parameters.   \\
$\Omega$ & RF drive frequency. \\
$\tau$ & Dimensionless unit of time, $\tau = \Omega t/2$. \\
$c_{2m,j}$ &  Fourier series coefficients of the Mathieu functions. \\
$\beta_j$ &  Mathieu characteristic exponent. \\
$W_j$ &   Wronskian, $W_j = \mathrm{ce}(a_j,q_j,0) \mathrm{\dot{se}}(a_j,q_j,0) $.  \\
$A_j$ & Amplitude of the intrinsic motion. \\
$\phi_j$ &  Phase of the secular motion.\\
$E_j$ & Secular energy. \\
$g_j(\tau)$ & External, spatially constant force. \\
$m_i , m_b$& Mass of the ion and buffer gas, respectively. \\
$\tilde{m} $& Mass ratio = $m_b/m_i$.   \\ 
$\dist{x}$& Probability distribution of the random variable $x.$ \\
$n_T$ & Tsallis (power-law) exponent. \\
$n_T^*$ & Estimate of $n_T$ from the multiplicative model. \\
$\hat{n_T}$ & Estimate of $n_T$ from the method of moments. \\
\hline
\hline
\end{tabular}
\end{table}

\subsubsection{Forced motion}
Experimentally, it is likely that the ion will experience additional forces apart from the trapping potential, requiring the introduction of corresponding terms in Eq.~\eqref{eq:mathieuHomogenous} \cite{berkeland98a}. The simplest case is when these forces are independent of the position of the ion but may depend on $\tau$.  Labelling the sum of these forces $g_j(\tau)$ and including this term in Eq.~\eqref{eq:mathieuHomogenous}  produces the inhomogenous Mathieu equation,
\begin{equation} \label{eq:mathieuInhomogenous}
\ddot r_j(\tau)  + [a_j - 2q_j \cos(2 \tau) ] r_j(\tau)= g_j(\tau) .
\end{equation}
The general solution to Eq.~\eqref{eq:mathieuInhomogenous} can be written in the form \cite{boyce17a}, 
\begin{equation}\label{eq:mathieuPositionInhomogenous}
r_j(\tau)   = r_{h,j}(A_j,\phi_j,\tau) + r_{f,j}(\tau) ,
\end{equation}
 which is the sum of the solution to the homogenous equation, i.e. Eq.~\eqref{eq:mathieuPositionHomogenous}, and a term $r_{f,j}(\tau) $ describing the response of the ion to the additional force.  We will refer to $r_{f,j}(\tau)$ as the ``forced motion'' of the ion by analogy to the forced motion of a harmonic oscillator. The forced motion is given by \cite{boyce17a},
\begin{equation} \label{eq:forcedMotion}
r_{f,j}(\tau) =  -\frac{\mathrm{ce_j}(\tau)}{W_j} \int \mathrm{se_j}(\tau) g_j(\tau) d\tau + \frac{\mathrm{se_j}(\tau)}{W_j} \int \mathrm{ce_j}(\tau) g_j(\tau),
\end{equation}
where   $W_j = \mathrm{ce_j}(0) \mathrm{\dot{se_j}}(0)$ is the Wronskian. In contrast to the intrinsic motion, this additional forced motion does not depend on $A_j,\phi_j$, and consequently,  the time-averaged kinetic energy of the forced motion may be orders of magnitude larger than the secular energy \cite{berkeland98a}. Throughout, we asume that $r_{f,j}(\tau)$ is a periodic function and remains bounded at all times. 

\begin{figure}[t]
\centering
\includegraphics[width=.9\linewidth]{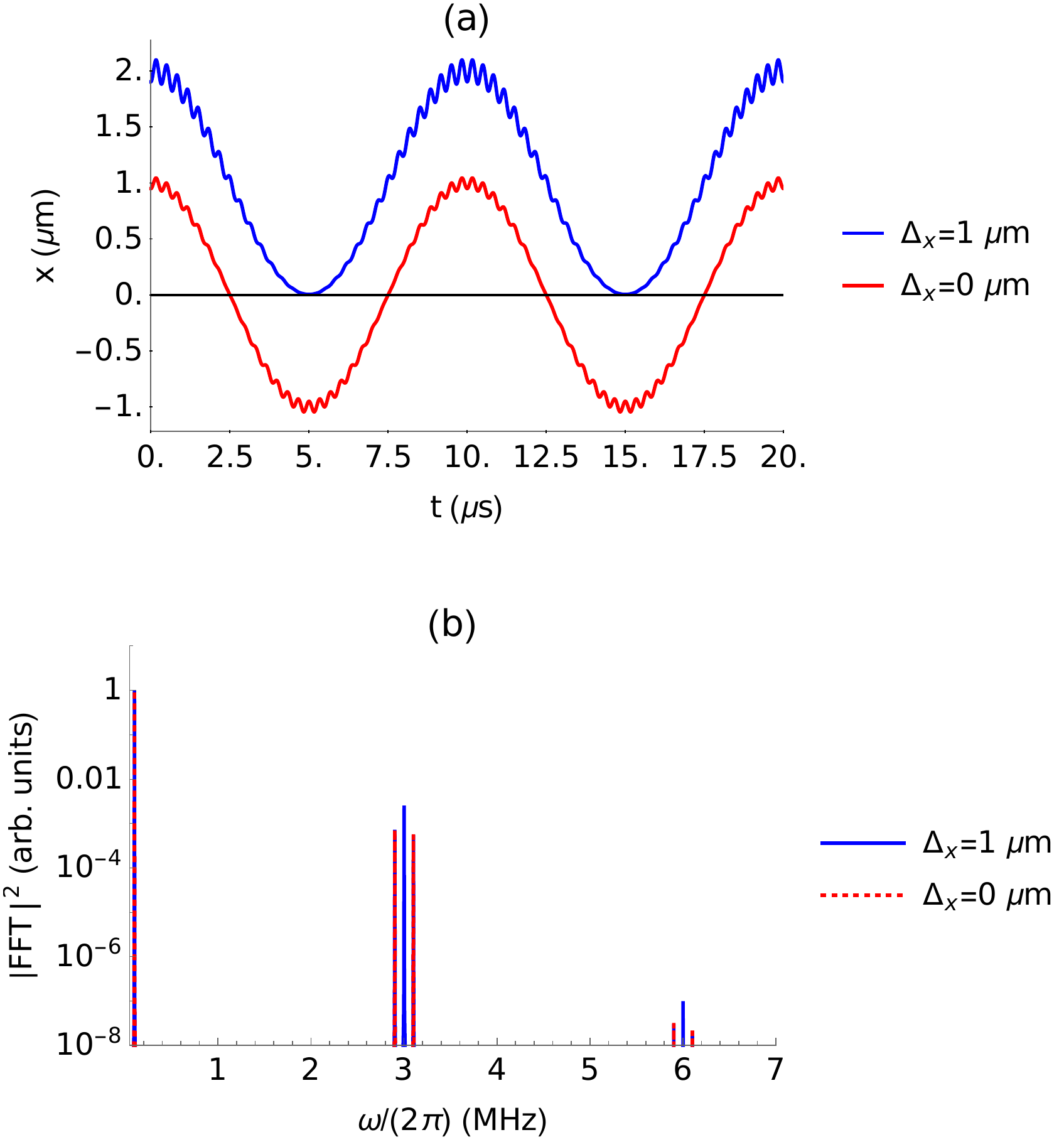}
\caption{(color online) (a) The trajectory of an ion in an rf trap with $\Omega = 3 \times 2 \pi$ MHz, $q_x = 0.1$  and $a_x$ defined such that the secular frequency is given by 100 $\times 2 \pi$ kHz.  The red (lower) curve shows the motion in the absence of an external force, wheras the blue  (upper) trajectory includes a static, spatially independent force  generating an offset of $\Delta_x = 1~\mu$m from the trap center. In both cases, the amplitude of the intrinsic motion is given by  $A_x = 1~\mu$m.   (b) The numerical Fourier transforms of the trajectories shown in (a). } \label{fig:trajectoryComponentsUniformField}
\end{figure}

The numerical results presented in this work employ a time-independent external force, $g_j(\tau) = g_j$, which for example represents the effects of a static uniform electric field \cite{berkeland98a}.  Substituting this into Eq.~\eqref{eq:forcedMotion} and evaluating the integrals produces,
\begin{equation} \label{eq:positionEMM}
r_{f,j}(\tau) =  \frac{g_j}{W_j}  \sum_l \sum_m  \frac{c_{2l,j} c_{2m,j}}{\beta_j + 2m} \cos[ 2(m-l)\tau] .
\end{equation}
The $m=l$ terms correspond to a constant offset of the ion's position while the $m\neq l$ terms are oscillations at multiples of $\Omega$. It is convenient to parameterise the effects of this force in terms of the displacement due to the $m=0,l=0$ term ,
\begin{equation} \label{eq:secularOffsetFromConst}
\Delta r_j = \frac{g_j}{\beta_j W_j} c_0^2 .
\end{equation}
If the force is due to an electric field $\mathbf{E}$ then,
\begin{equation}
g_j = \frac{1}{m_i} \frac{4}{\Omega^2} Q_i \mathbf{E} \cdot \hat{r_j} ,
\end{equation} 
where $Q_i$ is the charge of the ion and $\hat{r_j}$ is a unit vector. Substituting this into Eq.~\eqref{eq:secularOffsetFromConst} and expanding $W_j$ to first order in $m$, $W_j \approx c_{0,j}^2 \beta_j$, we find,
\begin{equation}
\Delta r_j = \frac{1}{m_i} \frac{4}{\Omega^2}  \frac{Q_i \mathbf{E} \cdot \hat{r_j}}{\beta_j^2 },
\end{equation}
in agreement with the result given in Ref.~\cite{berkeland98a}. The sum of the next two largest terms, $m=0,l=\pm 1$ give oscillations with a magnitude of approximately $\Delta r_j q_j/2$ at a frequency of $\Omega$ and are in phase with the rf trapping field.  This is distinct from the intrinsic micromotion, the components of which have amplitudes proportional to $A_j$, are sensitive to the phase of the secular motion $\phi_j$ and occur at frequencies offset from multiples of $\Omega$ by the secular frequency. For consistency with the literature, we will refer to this specific form of forced motion as ``in-phase excess micromotion'', but we stress that because this motion is independent of $A_j$ it is fundamentally different to the intrinisic micromotion, may well be orders of magnitude larger than the secular motion, and plays a different role during collisions. Fig.~\ref{fig:trajectoryComponentsUniformField} shows numerically simulated trajectories of the ion and their Fourier transforms for a fixed amplitude with and without EMM. 


\subsubsection{Collisions in the friction model}
As another example of the difference between IMM and EMM, and to motivate the rest of this work, we consider the effects of a damping force proportional to the velocity of the ion which may represent effects such as frequent collisions with atoms of a light buffer gas \cite{major05a}. The corresponding Mathieu equation is,
\begin{equation} \label{eq:mathieuDamped}
\ddot r_j + 2 \mu_j \dot r_j + (a_j - 2 q_j \cos 2\tau) r_j = g_j(\tau) ,
\end{equation}
where $\mu_j$ is the dimensionless form of the friction coefficient \cite{major05a}. The solution is (see Appendix~\ref{appendix:dampedMathieu}),
\begin{equation}
\begin{split}
r_j(\tau) =& A_j e^{-\mu_j \tau}[\mathrm{ce}(\tilde{a}_j  ,q_j,\tau) \cos \phi_j   -  \mathrm{se}(\tilde{a}_j  ,q_j,\tau) \sin \phi_j] \\ &+ r_{f,j}(\tau),
\end{split}
\end{equation}
where $\tilde{a}_j = a_j - \mu_j^2$, and $r_{f,j}(\tau)$ is found through the variation of parameters. The most significant result of the introduction of damping is that  the intrinsic motion exhibits an exponential decay towards zero. In contrast, the forced motion due to a constant or periodic $g(\tau)$ does not exhibit this decay, see Appendix~\ref{appendix:dampedMathieu}. This is another example of the difference between the IMM and forced motion -- if cooling is present, then the intrinisic micromotion eventually decays to zero wheras the forced motion does not.  

In reality, the amplitude of the intrinsic motion is prevented from reaching zero as a result of heating of the ion by recoil during collisions or the existence of other heating mechanisms. The energy transferred in each collision is a random quantity and so the secular energy of the ion is no longer a fixed quantity but must be treated as a random variable.  In the model of frequent collisions with light atoms and in the absence of forced motion, this leads to a Gaussian distribution for the position and velocity of the ion and hence $E_j$ follows a Boltzmann distribution, consistent with a particle in thermal equilibrium with a heat bath \cite{siemers88a}. In the opposite regime, in which collisions take place infrequently and with atoms of a non-negligible mass, the friction model considered above is no longer valid. Furthermore, the observed energy distributions are no longer adequately described by thermal statistics, but instead exhibit a power-law tail \cite{devoe09a,zipkes11a,chen14a,meir16a,rouse17a}. It is therefore necessary to investigate the effects of each collision in more detail to explain these results.

\subsection{Ion-neutral collisions}
To simplify the problem, it is assumed that collisions are classical, short range, and instantaneous such that the ion's trajectory is defined at all times by Eq.~\eqref{eq:mathieuInhomogenous}. The trajectory after the collision must therefore have the same general form as Eq.~\eqref{eq:mathieuPositionInhomogenous}, but with the constants of integration $A_j,\phi_j$ updated to new values,
\begin{equation}\label{eq:mathieuPositionInhomogenousPostCollision}
r'_j(\tau)   =r_{h,j}(A'_j,\phi'_j,\tau) + r_{f,j}(\tau) ,
\end{equation}
where primes indicate post-collision quantities. Note that since $r_{f,j}(\tau)$ does not depend on $A_j,\phi_j$ it is identical before and after the collision, wheras both the magnitude and phase of the intrinsic motion may be altered. For an instantaneous collision, the ion's position must remain unchanged. Equating $r'_j(\tau)$ and $r_j(\tau)$, then subtracting $r_{f,j}(\tau)$ from both sides produces,
\begin{equation} \label{eq:thermalPositionEq}
r_{h,j}(A'_j,\phi'_j,\tau) = r_{h,j}(A_j,\phi_j,\tau) .
\end{equation}
Next, we consider the velocity after a collision. We assume a model of elastic, hard-sphere collisions in which the post-collision velocity  is given by \cite{devoe09a, zipkes11a, chen14a, rouse17a} ,
\begin{equation} \label{eq:collisionAtomVel1}
\mathbf{v'}  =  \frac{1}{1 + \tilde{m}}  \mathbf{v}   +  \frac{\tilde{m}}{1 + \tilde{m}}  \mathbf{v_b}  + \frac{\tilde{m}}{1 + \tilde{m}}   \mathrm{R}  (\mathbf{v} -\mathbf{v_b} ) ,
\end{equation}
where bold-faced variables indicate vectors, e.g., $\mathbf{v} = (v_x,v_y,v_z)^T$, $\mathbf{v_b}$ is the velocity of the colliding particle of buffer gas, $\tilde{m} = m_b/m_i$ is the buffer gas-to-ion mass ratio, and $R$ is a rotation matrix determined by the scattering angles. As with the position, the velocity of the ion is given by the sum of the intrinsic and forced motion, $v_j(\tau) = v_{h,j}(A_j,\phi_j,\tau) + v_{f,j}(\tau)$, where the forced term is independent of $A_j,\phi_j$ and so is unchanged by the collision. We therefore obtain,
\begin{equation} \label{eq:collisionAtomVel2}
\mathbf{v_h'}   =  \frac{1}{1 + \tilde{m}}  \mathbf{v_h}   +  \frac{\tilde{m}}{1 + \tilde{m}}  (\mathbf{v_b}-\mathbf{v_f})  + \frac{\tilde{m}}{1 + \tilde{m}}   \mathrm{R} [\mathbf{v_h} -(\mathbf{v_b} - \mathbf{v_f})] .
\end{equation}
Taken together, Eq.~\eqref{eq:thermalPositionEq} and Eq.~\eqref{eq:collisionAtomVel2} indicate that the problem is equivalent to that of an ion with no forced motion colliding with a particle of velocity $\mathbf{v_b} - \mathbf{v_f}$. This is similar to the frame transformation used in Ref.~\cite{hoeltkemeier16a}  in which the intrinisic micromotion is assigned to the buffer gas, but in the present case is performed only for the forced motion and is valid for all $q_j$. 

Using the procedure detailed in Ref.~\cite{rouse17a}, we obtain a set of coupled equations for $A_j'^2$ and hence the secular energies after a collision,
\begin{equation} \label{eq:energyComponents}
\begin{split}
E'_j  = \sum_{(k,l) \in (x,y,z) } &\bigg (  \eta_{jkl} \sqrt{E_k} \sqrt{E_l} + a_{1,jkl} \sqrt{E_k} v_{b,l} \\ &+   a_{2,jkl} v_{b,k} v_{b,l}  +  a_{3,jkl} \sqrt{E_k} v_{f,l} \\ &+  a_{4,jkl} v_{f,k} v_{f,l}   +  a_{5,jkl} v_{f,k} v_{b,l} \bigg) ,
\end{split}
\end{equation}
where the coefficients $\eta_{jkl}$ and $a_{i,jkl}$ describe the transfer of energy between the motion along the three coordinate axes and between the different components of the ion's velocity and the velocity of the buffer gas. The coefficients of this expression depend on the elements of the random rotation matrix $\mathrm{R}$, the set of phases $\phi_j$, and the time of collision $\tau$. As supplementary information to this article, we provide a Mathematica notebook containing details of this procedure and the full form of Eq.~\eqref{eq:energyComponents}.  
%

To gain a better understanding of the collision process, it is useful to average over the collision parameters to obtain the mean post-collision energy for a given set of pre-collision energies, $\langle E'_j | E_x,E_y,E_z \rangle$. To do so, we must introduce some further assumptions. The Langevin model of collisions has been shown to be accurate for the classical trajectories considered here \cite{zipkes11a,chen14a} and so we adopt this. This results in two useful simplifications. Firstly, the rotation matrix $\mathrm{R}$ is isotropic in this model and so is uncorrelated from the velocity of the ion and neutral particle. Secondly, collisions occur at a uniform rate which is independent of the energy of the ion, and so $\tau$ can be assumed to follow a uniform distribution. We assume that the density of the buffer gas is low and uniformally distributed in space, which results in collisions occuring with equal probability at all points in the ion's trajectory, such that $\phi_j$ follows a uniform distribution. We also assume that the velocity of the buffer gas follows Maxwell-Boltzmann statistics and is characterised by a fixed temperature $T_b$. Both the density and the temperature of the buffer gas are taken to remain constant, i.e., the heating of the buffer gas due to the collisions is assumed to be negligible.  With these assumptions, the averaging can be performed over $\phi_j$, $v_{b,k}$, $\tau$ and the elements of the isotropic random rotation matrix by integrating the coefficients over the distributions of each of these variables in turn,  see the supplemental material for details. The coefficients of the terms in Eq.~\eqref{eq:energyComponents} where $k \neq l$ average to zero, as do the $c_1,c_3,c_5$ coefficients, significantly simplifying the expression. The remaining terms are given by,
\begin{equation} \label{eq:partialMeanEnergyComponents}
\langle E'_j | E_{x,y,z} \rangle  = \sum_{k \in (x,y,z) } \left[  \langle \eta_{jk} \rangle   E_k   +\langle  a_{4,jk} v_{f,k}^2   \rangle \right] + \kappa_j  k_B T_b  ,
\end{equation}
where the coefficients are defined as,
\begin{equation}
\langle \eta_{jk} \rangle = \frac{\delta _{jk}}{\tilde{m}+1}+ \frac{\tilde{m}  \kappa_j
  (3 \delta _{jk}+1) }{6   \beta_k^2 c_{0,k}^2  } \mathcal{M}_j  \left[
\mathrm{\dot{ce_k}}(\tau )^2+\mathrm{\dot{se_k}}(\tau )^2 \right ],
\end{equation}
and, 
\begin{equation} \label{eq:c4DrivenCoeff}
\langle  a_{4,jk} v_{f,k}^2  \rangle = \frac{ \tilde{m}  m_i \Omega^2  \kappa_j }{24  }(3 \delta _{jk}+1) \mathcal{M}_j\left[v_{f,k}(\tau)^2\right].
\end{equation}
In the above expressions,  $\delta_{jk}$ is the Kronecker delta, $\kappa_j$ is defined by,
\begin{equation}
\kappa_j = \frac{\tilde{m}}{ (1+\tilde{m})^2} \frac{\beta_j^2 c_{0,j}^2  }{ W_j^2},
\end{equation}
 and the operator $\mathcal{M}_j$ is defined as,
\begin{equation}
\mathcal{M}_j \left[ h(\tau)  \right] = \lim_{L\rightarrow \infty} \frac{1}{2L} \int^L_{-L}  h(\tau)   \left[\mathrm{ce_j}(\tau )^2+\mathrm{se_j}(\tau )^2\right]    d \tau.
\end{equation}
In principle, the above procedure may be adapted to arbitrary distributions for the velocity of the buffer gas by averaging over these in place of the Maxwell-Boltzmann distribution. This would allow for an investigation of the results when, e.g., Fermi-Dirac or Bose-Einstein statistics are required to correctly characterize the buffer gas. In practice, however, such statistics become relevant at collision energies low enough so that a classical description of the motion may no longer be valid. Moreover, at such low energies the long-range nature and finite duration of the ion-neutral interaction introduces an additional heating effect from dislocating the ion from its position in the rf field. At higher collision energies, this heating effect becomes less significant \cite{cetina12a}. To simplify both the analytical model and the numerical calculations, we proceed by assuming that the energy of the ion is large enough so that these effects can be neglected. 

The mean energy after a large number of collisions can be calculated from  Eq.~\eqref{eq:partialMeanEnergyComponents} as follows. Averaging over the pre-collision energies, corresponding to setting $\langle E'_j | E_x,E_y,E_z \rangle \rightarrow \langle E'_j \rangle$ and $E_j \rightarrow \langle E_j \rangle$, produces
\begin{equation} \label{eq:meanEnergyComponents}
\langle E'_j \rangle = \sum_k \langle \eta_{jk} \rangle \langle E_k \rangle + \langle \epsilon_j \rangle,
\end{equation}
where  $\langle \epsilon_j \rangle$ is the sum of the energy-independent terms arising in the averaging procedure. Depending on the values of the $\langle \eta_{jk} \rangle$, the mean energy will either increase with every collision, or tend towards a finite value for which $\langle E'_j \rangle = \langle E_j \rangle$. In the latter case, substituting this equality into Eq.~\eqref{eq:meanEnergyComponents} and solving for the mean energies produces,
\begin{equation} \label{eq:meanEnergyMatrix}
\begin{pmatrix}
\langle E_x \rangle \\
\langle E_y \rangle \\
\langle E_z \rangle
 \end{pmatrix} = 
\left( \mathrm{I_3} - 
\begin{pmatrix} 
\langle \eta_{xx} \rangle & \langle \eta_{xy}\rangle & \langle \eta_{xz} \rangle\\
\langle \eta_{yx} \rangle &\langle  \eta_{yy} \rangle& \langle \eta_{yz}\rangle\\
\langle \eta_{zx} \rangle & \langle \eta_{zy} \rangle& \langle \eta_{zz}\rangle
\end{pmatrix}
 \right)^{-1}
 \cdot
\begin{pmatrix} 
\langle \epsilon_x \rangle \\
\langle \epsilon_y \rangle \\
\langle \epsilon_z \rangle
\end{pmatrix},
\end{equation}
where  $\mathrm{I_3}$ is the $3 \times 3$ identity matrix.   The mean total kinetic energy, $\langle E_{j,K} \rangle$,  of the ion including the contributions from the micromotion and the forced motion can then be evaluated from the values of $\langle E_j \rangle$   (see Appendix~\ref{appendix:energyConversion}),
\begin{equation}
\langle E_{j,K} \rangle = \frac{ \langle E_j \rangle}{2 \beta_j^2 c_{0,j}^2}  \sum_m c_{2m}^2 (\beta + 2m)^2 + \frac{1}{2} m_i \frac{\Omega^2}{4} \overline{v_{f,j}^2} ,
\end{equation}
where $\overline{v_{f,j}^2}$ is the mean-square velocity of the forced motion.  For simplicity, however, we will continue to use the secular energy, since for low mass ratios and low values of $q_j$ this is approximately equal for each axis, whereas the time-averaged energy is significantly larger for axes with $q_j \neq 0 $ compared to axes with $q_j = 0$ \cite{chen14a}.  It is possible that the matrix inversion in Eq.~\eqref{eq:meanEnergyMatrix} cannot be performed or results in a negative energy, corresponding to a breakdown of the assumption that $\langle E'_j \rangle = \langle E_j \rangle$ and implying that the mean energy does not converge to a fixed value. The mass ratio at which this occurs for a given set of trapping parameters is referred to as the critical mass ratio \cite{chen14a}, and since it is independent of the $\epsilon_j$, it is unchanged when forced motion is present. 

\begin{figure}[t]
\centering
\includegraphics[width=.9\linewidth]{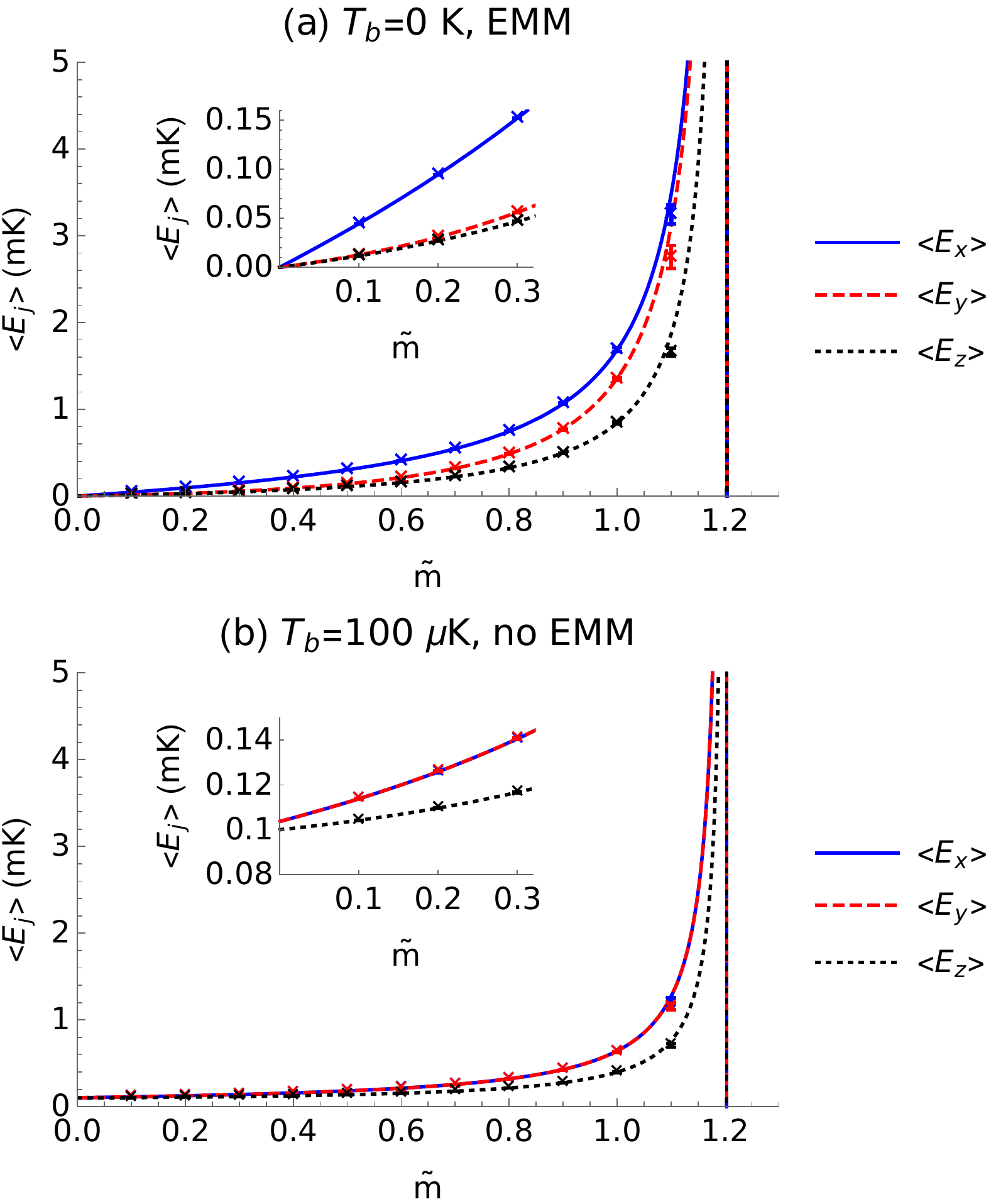}
\caption{(color online) (a) The analytically calculated value of the mean secular energy $\langle E_j\rangle$ for $j = x$ (blue solid line), $y$ (red dashed line) and $z$ (black dotted line) as a function of the neutral-to-ion  mass ratio, $\tilde{m}$, due to the presence of excess micromotion (EMM) and collisions with a buffer gas of temperature $T_b=0$~K. The points indicate the results obtained from numerical simulations ($10^6$ iterations for each value of $\tilde{m}$).  The trap parameters correspond to an ideal linear trap with $q_r= 0.2$, $a_z  = 0.000625$ and the excess micromotion is generated by a uniform electric field with a magnitude of $\approx 0.5$ V/m, corresponding to a displacement of the ion from the centre of the trap by $100$~nm along the $x$-axis. The inset shows the trend as $\tilde{m}\rightarrow 0$. Error bars represent the standard error of the mean calculated from the numerical data and are typically smaller than the size of the symbols. (b) As (a), but with a buffer gas temperature of $T_b= 100~\mu$K and no EMM.  In both figures, the vertical asymptote indicates the point at which the mean energy diverges, see main text for details. }
\label{fig:meanEnergyMassRatio}
\end{figure}

We now focus on the case of in-phase EMM in an ideal linear quadrupole trap defined by $q_r$ and $a_z$, taking $q_x = -q_y = q_r$, $q_z = 0$, and $a_x = a_y = -a_z/2$. The constant offset in the ion's position caused by the spatially-independent force does not appear in either Eq.~\eqref{eq:thermalPositionEq} or Eq.~\eqref{eq:collisionAtomVel2}, and so the most significant effect is the oscillations described by  $v_{f,j}(\tau) \propto \sin(2\tau)$.  Hence, the present results can also be adapted to the case of excess micromotion due to an rf phase offset, which also results in forced motion with the same form of the velocity \cite{berkeland98a}. In Fig.~\ref{fig:meanEnergyMassRatio}(a), the predicted mean secular energies for the case of excess micromotion along the $x$-axis and $T_b = 0$~K are shown and compared to the results of numerical simulations (see Appendix~\ref{appendix:numerical} for details), while the results in the absence of EMM but with a nonzero buffer-gas temperature ($T_b = 100~\mu$K) are shown in Fig.~\ref{fig:meanEnergyMassRatio}(b) for comparison.  At low mass ratio, there is a clear difference in the qualitative behaviour of the mean energies obtained for the two cases. In this regime, forced motion leads to one high-energy component ($E_x$) and two components with lower energy ($E_y, E_z$), and in the limit $\tilde{m} \rightarrow 0$ these all converge to $0$. In contrast, a nonzero value of $T_b$ results in two high-energy components and one low-energy component   which converge to non-zero values as $\tilde{m} \rightarrow 0$. Note that the radial and axial components have different mean values even in this limit. As the mass ratio increases, the transfer of energy between the motion along the $x$ and $y$ axes becomes more efficient such that in both cases there are two high-energy and one low-energy component. We may therefore predict that the differences between the two cases will be most significant at low mass ratio. In both cases, all energies diverge at the same mass ratio $\approx 1.2$, confirming that the critical mass ratio is unaffected by the presence of forced motion.


\section{Energy distributions}
It has previously been established that the distribution of the ion's energy does not, in general, follow a thermal distribution regardless of whether or not forced motion is present, and that it closely follows Tsallis statistics in both cases \cite{devoe09a,zipkes11a,chen14a,meir16a,rouse17a}. Before proceeding further, we must first confirm that the Tsallis distributions obtained through numerical simulations can be successfully predicted from our analytical model for the collision process. To simplify matters, we investigate the distribution of the total secular energy $E = E_x + E_y + E_z$ by making the change of basis,
\begin{equation}
\begin{split}
E_x &= E \sin^2 {\theta_\rho} \cos^2 {\phi_\rho}, \\
E_y &= E \sin^2 {\theta_\rho} \sin^2 {\phi_\rho}, \\
E_z &= E \cos^2 {\theta_\rho} ,
\end{split}
\end{equation}
where $\theta_\rho,\phi_\rho \in [0,\pi/2)$ describe the relative distribution of the total energy $E$ between the three axes. The advantage of this basis is that it allows $E$ to be factored out of expressions involving $\sqrt{E_k}\sqrt{E_l}$ in Eq.~\eqref{eq:energyComponents}, e.g.,
\begin{equation}
\eta_{jxy} \sqrt{E_x} \sqrt{E_y} = E \eta_{jxy} \sin^2 \theta_\rho \sin \phi_\rho \cos \phi_\rho .
\end{equation}
Summing over $j$ in Eq.~\eqref{eq:energyComponents}, applying this change of basis,  and neglecting terms with a mean value of zero we obtain,
\begin{equation} \label{eq:energyPrimeTotalApprox}
E' = \eta E  +  \sum_j\left[ a_{2,j} v_{n,j}     + \sum_k (a_{4,jk} v_{d,k}^2   ) \right]   = \eta E + \epsilon,
\end{equation}
where $\eta$ contains the $\eta_{jkl}$ multiplied by functions of $\theta_\rho, \phi_\rho$. As a consequence of the random rotation of the trajectory during a collision, these two angles evolve on a faster timescale than $E$. Therefore, the correlations between $E$ and $\theta_\rho,\phi_\rho$ can be neglected and these angles averaged over, resulting in the linear recurrence relation Eq.~\eqref{eq:energyPrimeTotalApprox} for a single variable. However,  since these angles reflect the distribution of energy between the axes, their mean values will differ depending on the presence and form of forced motion, leading to a change in the distribution of $\eta$. That is, while the $\eta_{jk}$ are independent of the form of the additive noise, $\eta$ is not. Note that, however, for any given collision $\eta$ and $\epsilon$ are approximately independent of each other and of $E$.

\subsection{The existence of the steady state}
We now move on to the question of finding the energy distribution of the ion  given that it evolves in each collision according to Eq.~\eqref{eq:energyPrimeTotalApprox}. Linear stochastic recurrence relations of this form have been widely studied \cite{champernowne53a,kesten73a,levy96a, sornette97a,buraczewski16a}  and so we summarise the relevant results. Firstly, if $T_b = g_j(\tau) = 0$ such that $\epsilon$ is always zero, then $E'  = \eta E$. Note that in this model, $E = 0$ represents a fixed point, i.e., an absorbing state, since for $E = 0$ and any value of $\eta$ the result of a collision is $E' = E = 0$. The energy after $n$ collisions is given by \cite{rouse17a}, 
\begin{equation} \label{eq:energyNoAdditive}
E_{(n)}=E_{(0)}  \prod_{i=1}^{i=n}  \eta_{(i)}.
\end{equation}
Here, we use the notation $x_{(n)}$ to indicate the value of the variable $x$ at collision number $n$. We assume that collisions are infrequent enough that there is no correlation between them, and so the $\eta_{(i)}$ are independent and identically distributed. By taking the logarithm  of both sides of Eq.~\eqref{eq:energyNoAdditive}, we find $\ln E_{(n)} =\ln E_{(0)} + \sum_n \ln \eta_{(i)} $, and so $\ln E_{(n)}$ undergoes a random walk with steps of size $\ln \eta$ \cite{sornette97a}. The long-term behaviour of $\ln E_{(n)}$ therefore depends on the sign of $\langle \ln \eta \rangle$ to determine in which direction this random walk is biased. If $\langle \ln \eta \rangle > 0$, then $\ln E_{(n)} \rightarrow \infty$ as $n \rightarrow \infty$. Conversely, if $\langle \ln \eta \rangle < 0$, then $\ln E_{(n)} \rightarrow -\infty$ in this limit, and so the ion's energy tends towards zero.  Using the terminology of Ref.~\cite{buraczewski16a}, we refer to the $\langle \ln \eta \rangle < 0$ situation as the contractive case and $\langle \ln \eta \rangle > 0$ as the divergent case. 

For large $n$, the product $\prod \eta_{(i)} = \Pi_{\eta,n}$ follows a log-normal distribution and the distribution of $E_{(n)}$ may be found by averaging over the initial conditions \cite{riley10a},
\begin{equation} \label{eq:logNormalEnergy}
\dist{E_{(n)}} =  \int^{\infty}_0 \frac{1}{E_{(0)} }f_{\Pi_{\eta,n}}(E_{(n)}/E_{(0) } )\dist{E_{(0)}} d E_{(0)} .
\end{equation}

Evaluating this integral requires specifying the initial energy distribution, see Ref.~\cite{rouse17a} for the result when $E_{(0)}$ follows a thermal distribution. The energy distribution obtained through this method does not converge to a steady-state as the number of collisions increases which is a known property of an unbounded purely multiplicative random process \cite{champernowne53a, sornette97a,mitzenmacher04a,buraczewski16a}.   In the contractive case, each collision on average reduces the energy of the ion no matter how small it may already be, while if $\langle \ln \eta \rangle > 0$ the energy increases on average in each collision. Establishing a steady-state distribution requires either that the energy is bounded from below in the contractive case, or bounded from above in the divergent case \cite{sornette97a}. For the model considered in this work, there is no upper bound on the energy and so we will not consider the divergent case further, although note that a non
-uniform buffer gas can introduce an upper bound \cite{hoeltkemeier16a}. There is, however, a lower bound if at least one of $g_j(\tau)$ or $T_b$ are nonzero, since if this is true, then $\epsilon$ may take a non-zero value. Consequently, if $E << \epsilon$, then after a collision $E' = \epsilon$ and so the convergence towards $E=0$ is interrupted. This applies if $\epsilon$ has a non-zero probability to take any non-zero value no matter how small the resulting value may be. This is a result of the fact that when $E \rightarrow 0$, it will eventually become smaller than any non-zero value of $\epsilon$. In terms of the random-walk analogy used in Ref.~\cite{sornette97a}, the presence of $\epsilon$ corresponds to the introduction of a barrier preventing the energy from reaching the absorbing state at $E = 0$, altering the boundary conditions of the problem and hence leading to a different distribution. The combination of the drift towards this barrier due to $\eta$ (in the contractive case) and the reflection from it leads to a steady-state energy distribution exhibiting a power-law tail \cite{sornette97a}.  In contrast, the tail of the distribution obtained from Eq.~\eqref{eq:logNormalEnergy} depends on the initial conditions of the ion and does not exhibit a power-law tail for an initially thermal distribution \cite{touchette05a,rouse17a}.

\begin{figure}[t]
\centering
\includegraphics[width=.9\linewidth]{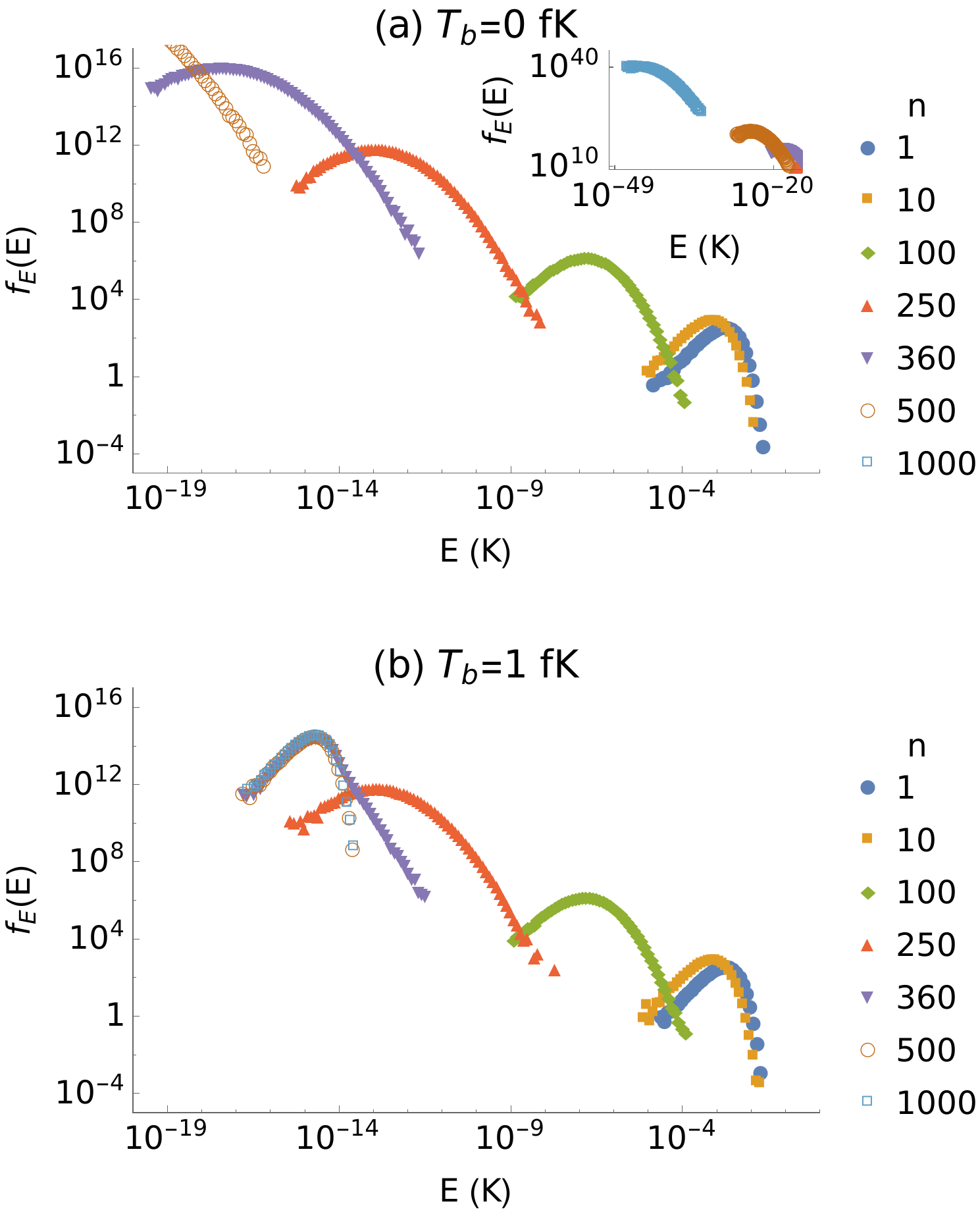}
\caption{(color online) The energy distribution of an ion in a linear rf trap with $q_r = 0.1, a_z=0.000625$, colliding with a buffer gas of neutral-to-ion mass ratio $\tilde{m} = 0.1$ after $n$ collisions. The ion's initial energy is taken from a thermal distribution with a temperature of $1~$mK, and the buffer gas temperature is set to either (a) $T_b = 0~$fK or (b) $T_b = 1~$fK. The inset in (a) shows the distributions obtained for $n=500$ and $n=1000$ collisions, which are not visible on the scale used for the main figure. 1'000'000 simulations are performed for each combination of collision number and $T_b$ to produce the numerical distributions.   }
\label{fig:collisionNumVarTemp}
\end{figure}

If the ion's initial energy is large compared to $\epsilon$, then it may take a large number of collisions for $E$ to reach the regime in which $\epsilon$ contributes significantly to the outcome of a collision. Consequently, for a small number of collisions the distribution may be close to the one obtained when $\epsilon$ is always zero \cite{sornette97a}. An order-of-magnitude estimate for the number of collisions  required for $\epsilon$ to become relevant to the dynamics may be found as follows. We denote this number of collisions $n_\epsilon$ and assume that $\langle \eta \rangle < 1$, that  $E_{(0)} >> \langle \epsilon \rangle$, and approximate that $\langle E_{(n)} \rangle \approx \langle \eta \rangle^n \langle E_{(0)} \rangle$. By setting $\langle E_{(n_\epsilon)} \rangle = \langle \epsilon \rangle$ we obtain,
\begin{equation} \label{eq:numCollisionsNeeded}
n_\epsilon = \frac{\ln( \langle \epsilon \rangle /\langle  E_0 \rangle)}{\ln (\langle \eta \rangle)} .
\end{equation}
As $\langle \epsilon \rangle \rightarrow 0$, the required number of collisions for the additive term to have an effect increases, but remains finite as long as $\langle \epsilon \rangle \neq 0$.  For typical trapping parameters $q = 0.1, a = 0.000625$, $\tilde{m} = 0.1$ and in the absence of excess micromotion, we find $\langle \epsilon \rangle \approx 0.25 k_B T_b$ and $\langle \eta \rangle \approx 0.92$ \cite{rouse17a}. Thus, for an ion with an initial temperature of $1~$mK, and a hypothetically very low buffer gas temperature of $T_b = 1~$fK, Eq.~\eqref{eq:numCollisionsNeeded} predicts that the ion's energy will be of the same order of magnitude as $\epsilon$ after approximately 360 collisions. This does not mean that the distribution has reached the steady-state by this point, but rather that $E$ is in the regime in which $\epsilon$ can no longer be neglected. In Fig.~\ref{fig:collisionNumVarTemp}, we plot the energy distributions obtained under these conditions for a varying number of collisions and compare these to the distributions obtained for the same parameters with $T_b = 0~$K. For the distributions corresponding to between $1$ and $250$ collisions, there is little difference between $T_b = 0$ and $T_b = 1~$fK, since the ion's energy is significantly larger than the additive term due to the temperature of the buffer gas. However, at greater collision numbers it can be seen that this is no longer the case, and a clear difference is visible at $360$ collisions, in agreement with the above prediction that this is when $\epsilon$ alters the dynamics. For $T_b = 0~$K,  the distribution continues to move towards lower and lower values of $E$ as the number of collisions increases, but for $T_b = 1~$fK the distributions for $500$ and $1000$ collisions are largely identical to each other, and are significantly different to the distributions obtained for the same number of collisions at $T_b = 0~$K. This is due to the influence of the lower bound on the energy caused by $\epsilon$, which in this case prevents $E$ from reaching values more than a few orders of magnitude lower than $10^{-15}$~K.  We reiterate that since $E$ otherwise decreases without limit, any non-zero value of $\epsilon$ is sufficient to produce a lower bound and a distribution with a power-law tail after a sufficiently large number of collisions, while if $\epsilon$ is always zero, then this lower bound does not exist and a qualitatively different distribution is obtained due to the change in boundary conditions. Although these two distributions are initially close (for the same initial conditions), they diverge as the number of collisions increases.  From this point on, we assume that at least one of $T_b$ or $g_j(\tau)$ are non-zero, and that the ion's energy distribution has reached the steady state.

 
\begin{figure}[t]
\centering
\includegraphics[width=.94\linewidth]{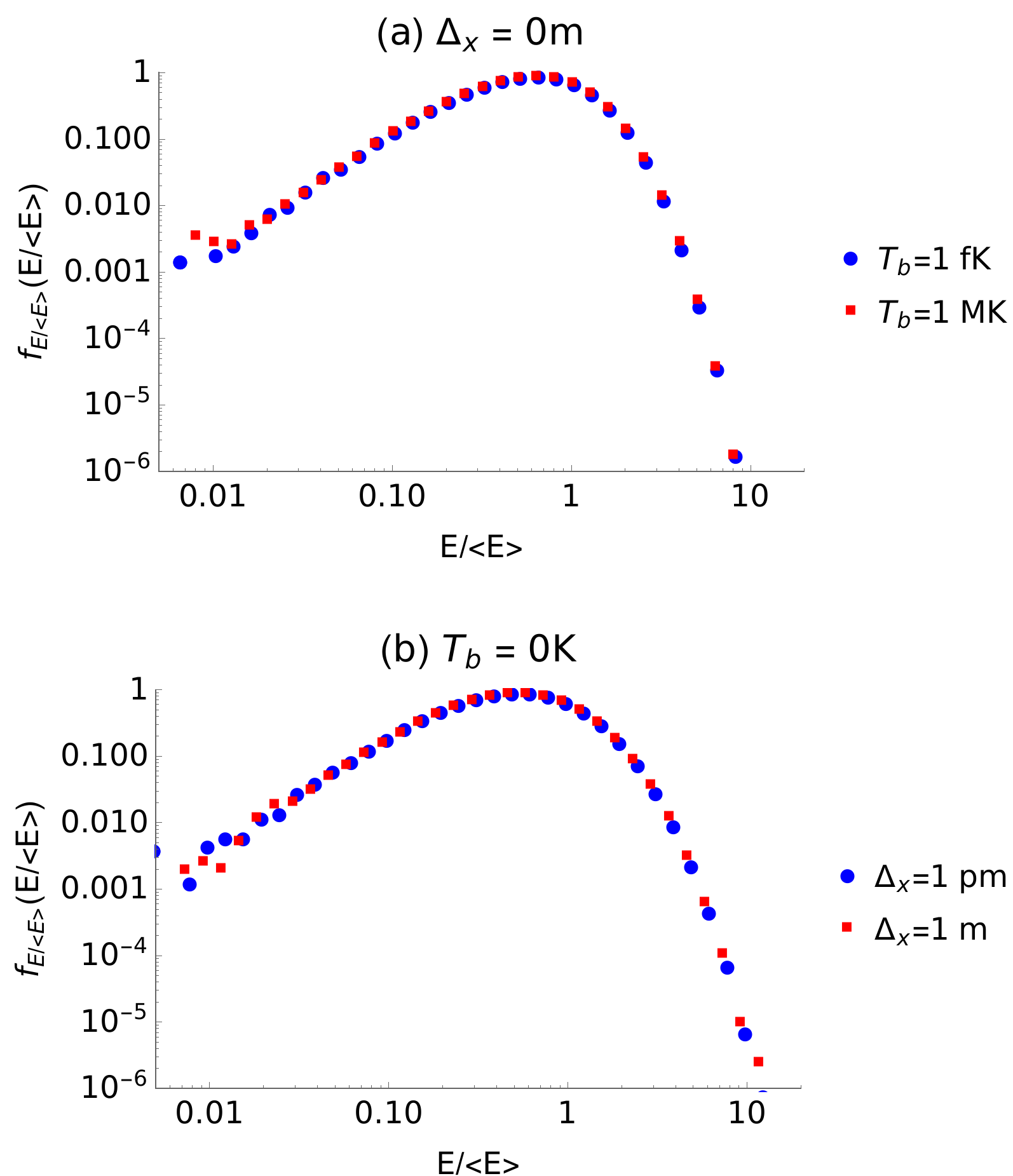}
\caption{(color online)  (a) The energy distribution of an ion in a rf trap with $q = 0.1, a=0.000625$, colliding with a buffer gas of neutral-to-ion mass ratio $\tilde{m} = 0.1$ after 1000 collisions with a buffer gas of temperature $T_b =1~$fK (blue circles) or $T_b = 1~$MK (red squares) in the absence of excess micromotion. (b) As (a), but with $T_b = 0~$K and excess micromotion parameterised by an offset of either $\Delta_x = 1~$pm (blue circles) or $\Delta_x = 1~$m (red squares) along the x-axis. The energies obtained have been rescaled by the mean energy for that distribution to make the similarity between the two distributions more apparent.  }
\label{fig:energy_dists_scaled}
\end{figure}

The form of the energy distribution does not depend on the units of energy apart from a constant scaling factor. That is, if the energy follows a distribution $\dist{E}$ and we define $\tilde{E} = a E$ where $a$ is a positive constant, then the distribution of $\tilde{E}$ is given by \cite{riley10a},
\begin{equation} \label{eq:randomDistScale}
\dist{\tilde{E}}  = \frac{1}{a} f_E(\tilde{E}/a).
\end{equation}
Since $\epsilon$ also has units of energy, it follows that we may choose these units such that a non-zero value of $\epsilon$ has an arbitrary magnitude without altering $\dist{E}$ beyond applying this scaling transformation. This means that multiplying $\epsilon$ by a fixed constant  is equivalent to changing the units of energy and therefore effectively applies a scaling factor to $\dist{E}$.  This property is why the magnitude of $\epsilon$ is unimportant in establishing the steady state, since we may always define units of energy in which $\epsilon$ is large, and it is reasonable to assume that the existence of the steady state does not depend on the units in which the energy is measured. The exception is if $\epsilon = 0$ in all cases, since then it will not be non-zero in any units of measurement. A particularly useful choice is to measure the energy in units of the mean energy, that is, taking $a =1/\langle E \rangle$, assuming that this exists and is not equal to zero. Doing so, we find that if $g_j(\tau) = 0$, then the same  distribution for $E/\langle E \rangle$ is obtained for any non-zero value of $T_b$,  see Fig.~\ref{fig:energy_dists_scaled}(a) for a comparison of $T_b = 1~$fK and $T_b = 1~$MK.   Likewise, the same result is obtained when setting $T_b = 0~$K and varying the amount of EMM, see Fig.~\ref{fig:energy_dists_scaled}(b) for offsets of $1~$pm and $1$m.  Note, however, that if both $T_b$ and $g_j(\tau)$ are nonzero simultaneously then rescaling one does not have the same effect, which we will discuss in more detail later. 


\subsection{Tsallis statistics}
The exact form of the steady-state energy distribution $\dist{E}$ depends on the distributions $\dist{\epsilon}$ and $\dist{\eta}$ but can be approximated by Tsallis statistics, that is, Eq.~\eqref{eq:tsallisDist}, when the heating is due to a nonzero value of $T_b$.  This result was derived in our previous work, Ref.~\cite{rouse17a}, by employing the formalism of superstatistics, in which the energy distribution is taken to be of the form \cite{beck03a},
\begin{equation} \label{eq:energyDistBeck}
\dist{E} = \int^{\infty}_{0} E^k \frac{1}{\Gamma(k+1)} \frac{1}{(k_B T)^{k+1}} e^{-E/(k_B T)} \dist{T}  d T.
\end{equation}
Eq.~\eqref{eq:energyDistBeck} expresses the energy distribution $\dist{E}$ as a thermal distribution averaged over the steady-state probability distribution of the secular temperature, $T$, and is related to the Laplace transform of the distribution of the inverse temperature. The value of $T$ is altered in each collision according to,
\begin{equation} \label{eq:temperatureMultiplicative}
T' = \eta T + \frac{\langle \epsilon \rangle}{(k+1) k_B},
\end{equation}
where the assumption has been made that the fluctuations in $E$ due to $\epsilon$ lead to an approximately constant increase in the temperature with each collision \cite{rouse17a}. That is, the variance of the additive term in the temperature domain is assumed to be negligible. This does not require that the additive term itself is small, only that the inverse Laplace transform of $\epsilon$ has a very narrow distribution. The steady-state form of $\dist{T}$ corresponding to Eq.~\eqref{eq:temperatureMultiplicative} is approximately given by an inverse Gamma distribution, and evaluating Eq.~\eqref{eq:energyDistBeck} using this distribution produces Eq.~\eqref{eq:tsallisDist} \cite{beck03a, biro05a, rouse17a}.  This distribution is defined by three parameters,   $\langle \beta \rangle, k$ and $n_T$. Of these, $k$ and $n_T$ are dimensionless, while $\langle \beta \rangle$ has units of inverse energy and so may be set to an arbitrary value by redefining the units of energy. That is, substituting the Tsallis distribution into Eq.~\eqref{eq:randomDistScale} produces,
\begin{equation}
\dist{\tilde{E}} =  \left(\frac{n_T}{\langle\beta\rangle}\right)^{-k-1}  \frac{\Gamma (k+n_T+1)}{\Gamma (k+1) \Gamma (n_T)}     \frac{\frac{1}{a}  (\tilde{E} /a)^k}{ \left(\frac{\langle\beta\rangle }{  n_T}\frac{\tilde{E}}{a}+1\right)^{k + n_T + 1}},
\end{equation}
and the factor of $a$ may be absorbed by defining $\langle \tilde{\beta} \rangle = \langle \beta \rangle / a$. Note that $n_T$ and $k$ are left unchanged by this rescaling, and so the overall shape of the distribution is unchanged as can be seen in Fig.~\ref{fig:energy_dists_scaled}.

The value of $k$ depends on the effective density of states. In the ideal case, this is simply the density of states for a three dimensional harmonic oscillator, leading to $k=2$. However, as noted in the previous section the mean energy for each axis differs such that not all degrees of freedom are equal. In the extreme case when the energy of one axis is much greater than the others, e.g., $E_x >> E_y, E_z$, then $E \approx E_x$, and so is approximately a one-dimensional system. Hence, the density of states would be much closer to that expected for a one dimensional harmonic oscillator, $k=0$. In practice, this effect is sufficiently small that we will simply assume that $k=2$ except for the purposes of fitting the Tsallis distribution to numerical data, for which $k$ is treated as a free parameter (Appendix.~\ref{appendix:numerical}). Thus, all that remains is to predict the value of $n_T$.

\subsection{Estimation of the Tsallis exponent}
If $T$ follows a linear stochastic recurrence relation with a constant, non-zero additive term, i.e., Eq.~\eqref{eq:temperatureMultiplicative}, and if $\eta$ has some probability of being greater than $1$, then the tail of $\dist{T}$ follows a power-law of the form $T^{-(\nu+1)}$, where $\nu$ is defined by $\langle \eta^\nu\rangle = 1$ \cite{sornette97a, buraczewski16a}.  If $\dist{T}$ exhibits a power-law tail, then, by the properties of the Laplace transform, $\dist{E}$ has the same power-law tail, which for Tsallis statistics is $\dist{E} \sim E^{-(n_T + 1)}$ for large $E$ \cite{touchette05a}. In Ref.~\cite{rouse17a}, we obtained an estimate for $n_T$, which we here denote $n_T^*$, by requiring that it satisfies  $\langle \eta^{n_T^* }\rangle = 1$. In this treatment the exponent is solely defined by the properties of the distribution of the multiplicative noise $\eta$. However, as discussed previously, the distribution of $\eta$ and hence the value of $n_T^*$ depends on $\phi_\rho,\theta_\rho$, which describe the distribution of the energy between the axes of motion. That is, if more of the total energy $E$ is associated with the motion along an axis with a large value of $q_j$, then the value of $\eta$ will typically be larger than if most of the energy is along an axis with a value of $q_j$ close to zero due to the greater amount of intrinsic micromotion in the first case. When the heating is due to a non-zero value of $T_b$, this  leads to a small but non-negligible effect on $n_T^*$, see Ref.~\cite{rouse17a}. Forced motion leads to a much greater change in how the energy is distributed between the axes of motion, as was shown in the previous section. Therefore, even if Eq.~\eqref{eq:temperatureMultiplicative} continues to accurately model the dynamics, we expect some difference in the value of $n_T^*$ obtained when forced motion is present as a result of the increase in the energy of one axis relative to the others. 

In general, if $T$ is a random variable then the resulting superstatistical energy distribution may be approximated by Tsallis statistics, even if there is no multiplicative noise \cite{beck03a, tsallis09a}. Moreover, if $\epsilon$ has a heavier tail than $\eta$, then the power-law tail of $E$ is defined from $\dist{\epsilon}$ and not $\dist{\eta}$ \cite{buraczewski16a}. Thus, since $n_T^*$ is calculated from $\dist{\eta}$, it may produce an incorrect estimate for the power-law tail and hence for $n_T$ if the additive fluctuations are larger than the multiplicative fluctuations.  We therefore introduce another estimator for $n_T$ by matching the moments of the Tsallis distribution to the analytical mean and mean-square energy, which does not require the assumption that the deviation from a thermal distribution is caused by  the multiplicative noise. The mean value of the Tsallis distribution is given by,
\begin{equation}
\langle E_T \rangle = \frac{(1+k)}{\langle \beta \rangle} \frac{n_T}{n_T -1} ,
\end{equation}
for $n_T > 1$.  The mean energy in terms of the collision parameters may be calculated using Eq.~\eqref{eq:meanEnergyMatrix}, and by equating $\langle E_T \rangle = \langle E \rangle = \langle E_x \rangle + \langle E_y \rangle + \langle E_z \rangle$ we obtain an equation relating $n_T$ to the mean energy. We require a second equation to eliminate $\langle \beta \rangle$, which is obtained from calculating the second moment of the Tsallis distribution $\langle E_T^2 \rangle$, and equating this to   $\langle E^2 \rangle = \sum_j \sum_k  \langle E_j E_k \rangle , (j,k) \in (x,y,z)$.  The $\langle E_j E_k \rangle$ are found by multiplying together $E'_j$ and  $E'_k$ as given by Eq.~\eqref{eq:energyComponents}, averaging over all the collision parameters and solving for the steady state, analogously to the mean energy. These second-order moments diverge at a lower mass ratio than the first-order moments, and in terms of the Tsallis distribution are defined only for $n_T > 2$. This requires small values of $\tilde{m}$ and $q_j$ and so we primarily focus on this regime from this point onwards. In terms of these mean energies, we find,
\begin{equation} \label{eq:ntFromEnergy}
\hat{n_T} = \frac{ (2+k) \langle E\rangle^2 - 2(1+k) \langle E^2 \rangle}{(2+k) \langle E \rangle^2 - (1+k) \langle E^2 \rangle} ,
\end{equation}
where $\hat{n_T}$ indicates that this is an estimation and is exact only if the distribution exactly follows Tsallis statistics with a known value of $k$, which following the discussion in the previous section we assume is given by $k=2$.  If the value of $\hat{n_T}$ is in good agreement with $n_T^*$ then we may take this as evidence that the  power-law tail is caused primarily by the multiplicative noise. However, if these estimates do not agree, then this indicates that another source of noise must be responsible for the deviation from thermal statistics.


\begin{figure}[t]
\centering
\includegraphics[width=.9\linewidth]{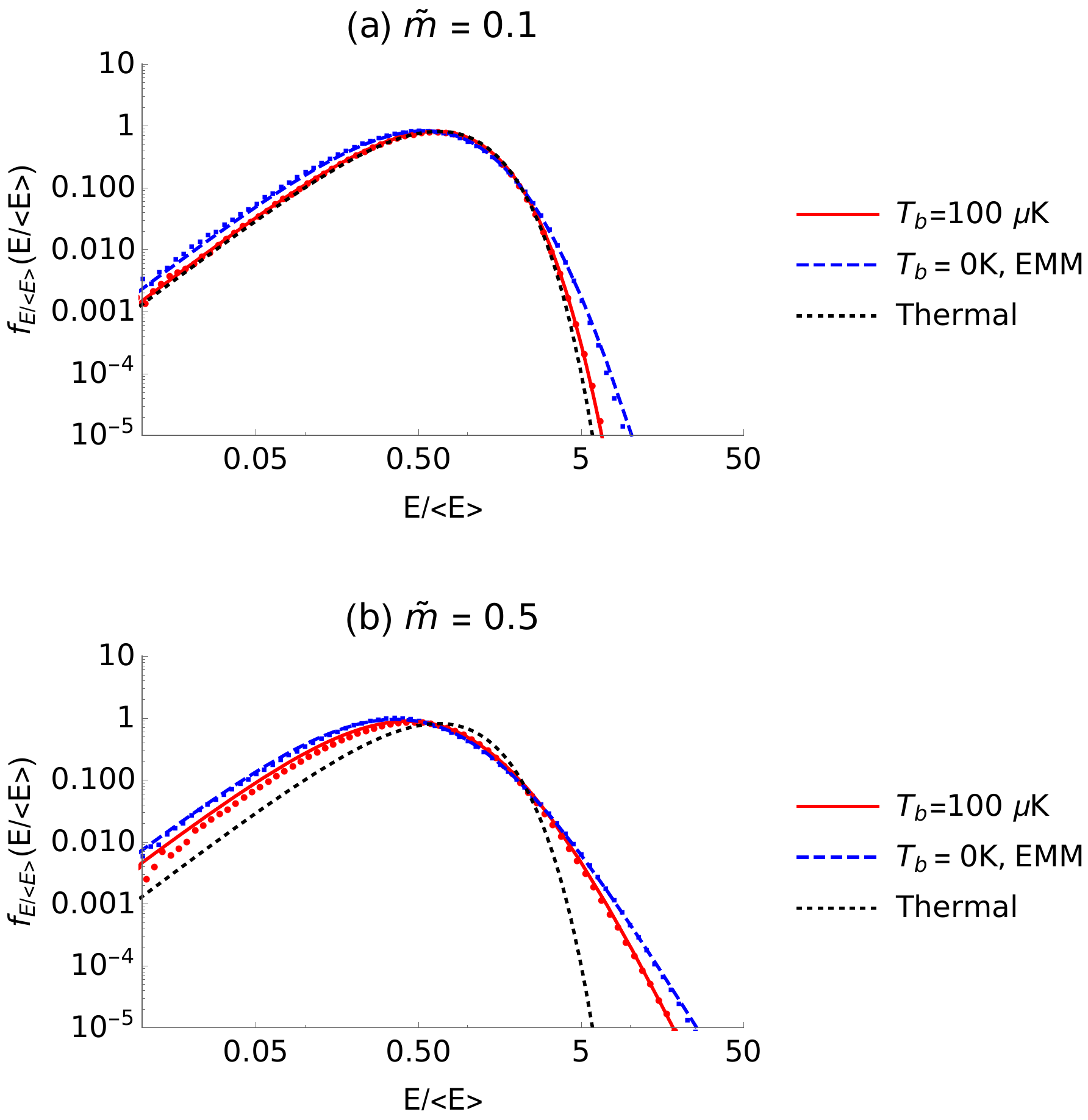}
\caption{(color online) The energy distribution for an ion exhibiting excess micromotion colliding with a buffer gas of temperature $T_b = 0$~K (blue, solid) and without excess micromotion colliding with a buffer gas of temperature $T_b = 100~\mu$K (red, dashed) for a) $\tilde{m}=m_b/m_i = 0.1$ and b) $\tilde{m}=m_b/m_i = 0.5$. The data have been scaled by the analytically calculated mean energy to make the difference between the two distributions more apparent. The trapping parameters are given by $q_r= 0.2,a_z = 0.000625$, and when present the excess micromotion is defined by a static electric field such that the equilibrium position is displaced by $100~$nm along the $x$-axis. The solid lines indicate the predicted Tsallis distributions while the dotted line gives the distribution for an ion in thermal equilibrium. Each distribution is obtained from 10'000'000 iterations of the numerical simulation and binned into logarithmically spaced bins, normalised by the bin width.}
\label{fig:thermalMicromotionCompare}
\end{figure}

To confirm that the use of Tsallis distributions and the values of $\hat{n_T}$ from Eq.~\eqref{eq:ntFromEnergy} are accurate, the distributions obtained from numerical simulations are compared to the distribution predicted using $\hat{n_T}$  for  $\tilde{m} = 0.1$ (Fig.~\ref{fig:thermalMicromotionCompare}(a))  and $\tilde{m}=0.5$ (Fig.~\ref{fig:thermalMicromotionCompare}(b)), finding good agreement.   At low mass ratio $\tilde{m} \approx 0.1$ and for the trapping parameters employed ($q_r = 0.2, a_z = 0.000625$ ),  it is generally assumed that the ion will exhibit a thermal energy distribution. It can be seen in Fig.~\ref{fig:thermalMicromotionCompare}(a) that this is approximately true in the absence of forced motion, for which the numerical data and predicted Tsallis distribution are both close to a thermal distribution. However, this does not hold when there is forced motion. The distribution still closely follows Tsallis statistics, but with a more pronounced power-law tail, i.e., a smaller value of $n_T$. As the mass ratio increases, the distribution for non-zero $T_b$ also deviates from a thermal distribution as expected, see Fig.~\ref{fig:thermalMicromotionCompare}(b). At high energies, a small deviation from the Tsallis distribution can be seen, typically accounting for $0.1\%$ of the data set. This is likely a result of the approximations made during the derivation of Tsallis statistics in Ref.~\cite{rouse17a} and the assumption that $k=2$.  Nonetheless, the bulk of the distribution is adequately described by the present treatment, and it is clear that there is a difference between the two cases.

\begin{figure}[t]
\centering
\includegraphics[width=.9\linewidth]{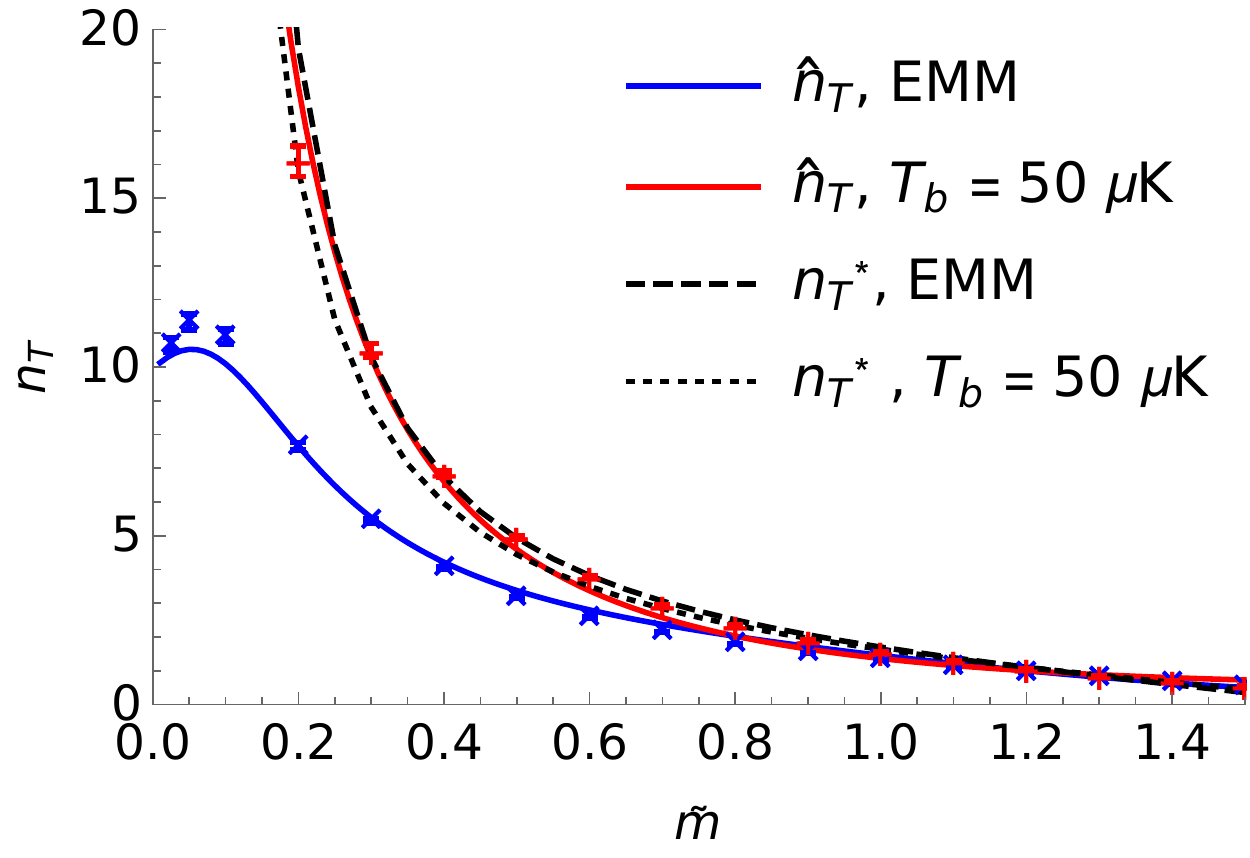}
\caption{(color online) A comparison of the analytically estimated  and numerically simulated values of the Tsallis exponent $n_T$ as a function of mass ratio for a   buffer gas with $T_b = 0~$K and a static electric field resulting in an offset of $100$~nm (blue, lower) and a buffer gas with a temperature of $T_b=50~\mu$K with no offset (red). The data points give the values found from maximum-likelihood estimation performed on the numerical data. The error bars indicate the calculated standard error and are typically smaller than the size of the symbols. The blue (lower) and red solid lines show the predicted value of the exponent from the analytically calculated mean and mean-square energy, denoted $\hat{n_T}$ in the main text.  The dashed and dotted lines show the prediction from the multiplicative coefficient $\eta$, $n_T^*$, for the thermal and forced cases, respectively \cite{rouse17a}. The trap parameters are given by $q_r = 0.1, a_z = 0.000625$ and 200'000 simulations are performed for each data point.  }
\label{fig:exponent_emm_ta}
\end{figure}

In Fig.~\ref{fig:exponent_emm_ta} we compare the exponents obtained from numerical simulations to both the predicted value due to multiplicative fluctuations from Ref.~\cite{rouse17a}, $n_T^*$, and the predicted value from Eq.~\eqref{eq:ntFromEnergy}, $\hat{n_T}$, as a function of mass ratio, both including and excluding EMM.  It can be seen that $n_T^*$ is a good predictor for the observed exponent in the absence of forced motion, as was demonstrated in Ref.~\cite{rouse17a}. Furthermore, at high mass ratio it also successfully predicts the exponent when forced motion is present, which is found to approach the value in the absence of forced motion. However, at low mass ratio there is no longer an agreement between $n_T$ and $n_T^*$, demonstrating that the  multiplicative model with an additive constant developed in Ref.~\cite{rouse17a} does not fully explain the dynamics in the regime of a low mass ratio with forced motion. In contrast, $\hat{n_T}$ remains reasonably accurate over all mass ratios. Both $\hat{n_T}$ and the numerical simulations show that  at low mass ratio the Tsallis exponent does not diverge to infinity if the ion is subject to forced motion, i.e. a thermal distribution is not obtained in this case.

\subsection{Additive fluctuations due to forced motion}
\begin{figure}[t]
\centering
\includegraphics[width=.9\linewidth]{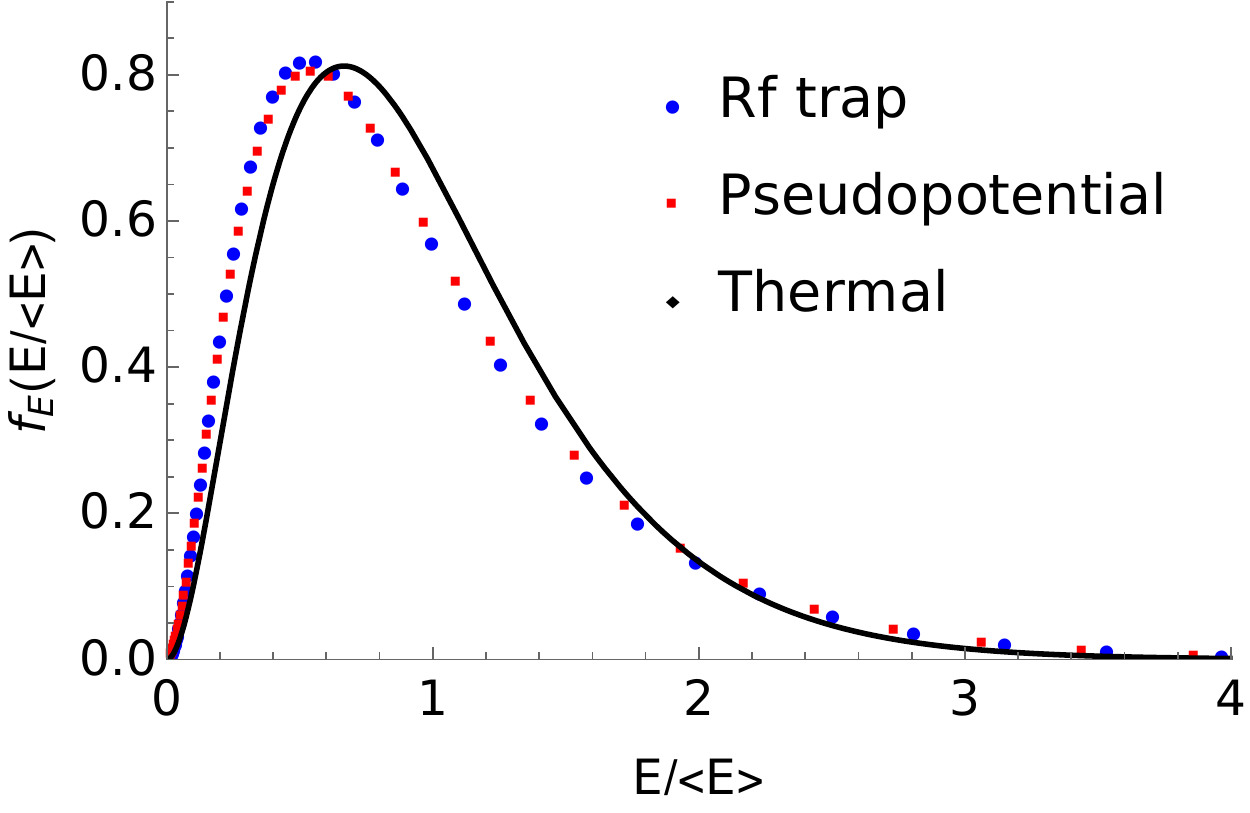}
\caption{(color online) The energy distribution  obtained for an ion in a linear rf trap ($q_r = 0.1, a_z = 0.000625$) interacting with a buffer gas of mass ratio $\tilde{m} = m_b/m_i = 0.1$ in a time-dependent trapping potential with excess micromotion corresponding to an offset of 100~nm along the $x$-axis (blue circles), compared to the distribution obtained for an ion in a harmonic  pseudopotential colliding with atoms with a velocity given by $v_x = \sin(2\tau)$ (red squares). The frequencies of the pseudopotential trap are set equal to the secular frequencies of the rf trap. The solid line indicates the distribution for a three-dimensional harmonic oscillator at thermal equilibrium. }
\label{fig:energy_forced_harmonic}
\end{figure}

The discrepancy at low mass ratio between the value of $n_T$ obtained from numerical simulations compared to the value estimated from $\dist{\eta}$, $n_T^*$, implies that the multiplicative fluctuations due to the micromotion interruption are not the only cause of the deviation from thermal statistics when forced motion is present. Thus, another source of fluctuations in the temperature must have an influence on $n_T$. We therefore re-examine the assumption in Ref.~\cite{rouse17a} that the additive fluctuations lead to a fixed increase in the temperature with each collision. In Eq.~\eqref{eq:collisionAtomVel2}, it is demonstrated that the velocity of the forced motion may be assigned to the buffer gas, but there is an important distinction between the thermal motion of the buffer gas and the forced motion in that the latter  does not follow a thermal distribution. To lowest order, the velocity of the in-phase EMM, i.e., the derivative of Eq.~\eqref{eq:positionEMM} with respect to $\tau$, is described by $v_{f,j}(\tau) = |v| \sin (2 \tau)$. When sampled at random collision times, $v_{f,j}^2$ follows a bimodal distribution with peaks of equal height at $0, |v|^2$, in contrast to the single peak for a thermal distribution \cite{arnold80a}. To demonstrate the importance of this, we perform simulations of an ion in a time-independent harmonic trap, i.e., in the pseudopotential approximation, undergoing collisions with a buffer gas with a velocity given by $v = |v| \sin (2\tau)$. This leads to the results shown in Fig.~\ref{fig:energy_forced_harmonic}, with the distribution close to that found when forced motion is present in an rf trap. Thus, a non-thermal velocity distribution of the buffer gas is sufficient to cause the deviation from thermal statistics for the ion even in the absence of the time-dependent trapping potential. 

As a toy model to better understand this situation, we assume that each collision samples one of the two peaks of the distribution of $v_f^2$ as if the ion had collided with a buffer gas of temperature either $0$ or $T_b$ with equal probability. The temperature then evolves according to,
\begin{equation} \label{eq:tempDichot}
T' = \eta T + B \langle \epsilon \rangle/(3 k_B),
\end{equation}
where $B$ takes values of $0$ or $1$ with equal probability, and $\epsilon$ is defined as for a thermal buffer gas with temperature $T_b$. In this model, we may view the temperature of the ion as being subject to dichotomous noise in addition to the multiplicative noise, leading to a different distribution than the one obtained for a fixed atomic temperature \cite{sancho84a,dubkov03a}. However, as shown in Ref.~\cite{beck03a}, the energy distribution obtained will still  approach Tsallis statistics as long as the ion's energy remains low. 

\begin{figure}[t]
\centering
\includegraphics[width=.92\linewidth]{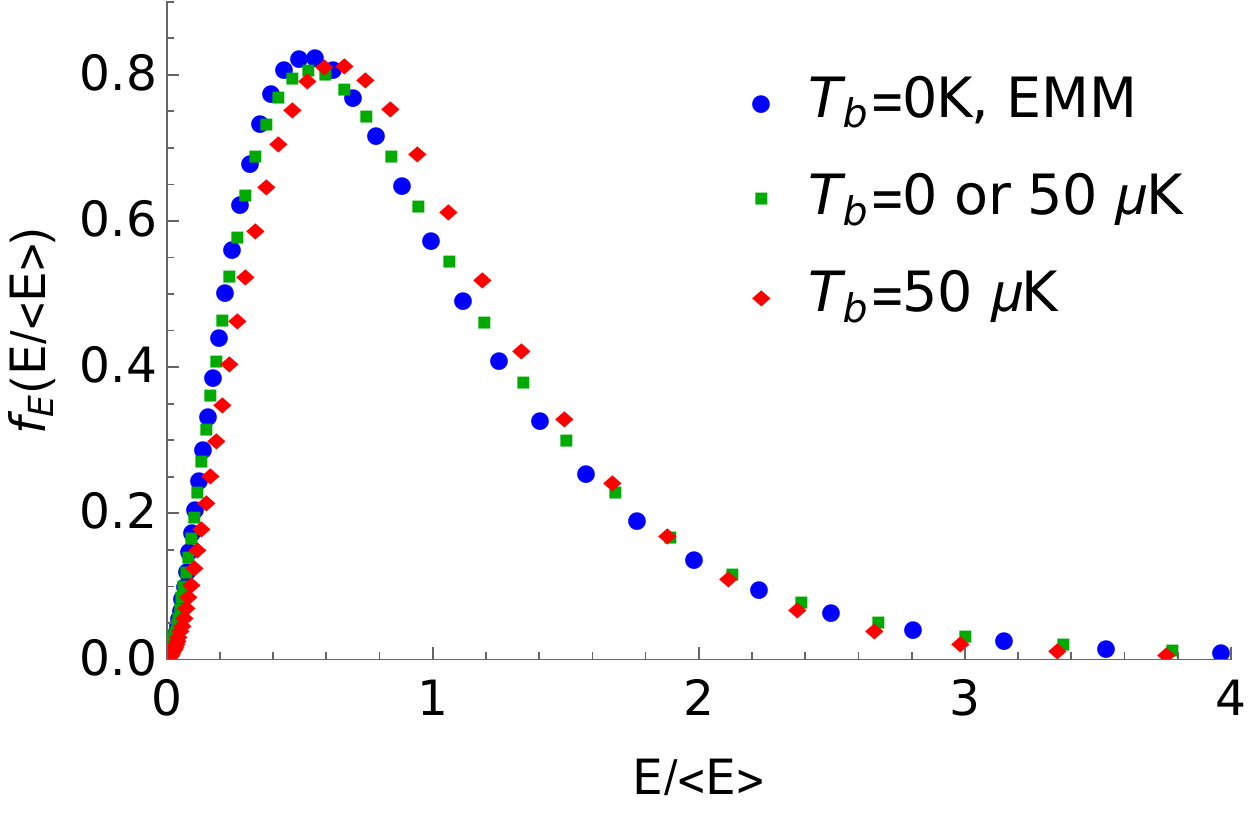}
\caption{(color online) A comparison of the energy distributions obtained for an ion in a linear quadrupole trap defined by $q_r = 0.1, a_z = 0.000625$ interacting with a buffer gas of mass ratio $\tilde{m}=m_b/m_i = 0.1$. Three cases are illustrated: forced motion and a buffer gas temperature of $T_b=0~$K (blue circles), a fixed buffer gas temperature of $T_b=50~\mu$K (red diamonds), and a buffer gas temperature which is randomly chosen in each collision from either $T_b=0$ or $50~\mu$K with equal probability (green squares). 10'000'000 simulations are performed for each of the three cases. }
\label{fig:energy_forced_dichot}
\end{figure}

To test this interpretation, in Fig.~\ref{fig:energy_forced_dichot} we show the energy distribution for a simulation in which the atomic temperature is chosen from either $T_b = 0$ or $50~\mu$K with each collision, which produces a distribution close to that observed in the presence of forced motion and which is noticeably different to the one obtained for the same trapping parameters with a fixed buffer gas temperature. It is interesting to compare this to the system discussed in Ref.~\cite{rouse15a}, in which an ensemble of ions underwent a combination of laser cooling with rare collisions with background gas leading to a large amount of heating. Neglecting the heating due to photon recoil, this situation is equivalent to Eq.~\eqref{eq:tempDichot} with   $B$ biased such that it has only a small probability of taking the value $1$ and $\eta$ fixed to a constant, which we demonstrated leads to Tsallis statistics \cite{rouse15a}.  
 
\begin{figure}[t]
\centering
\includegraphics[width=.9\linewidth]{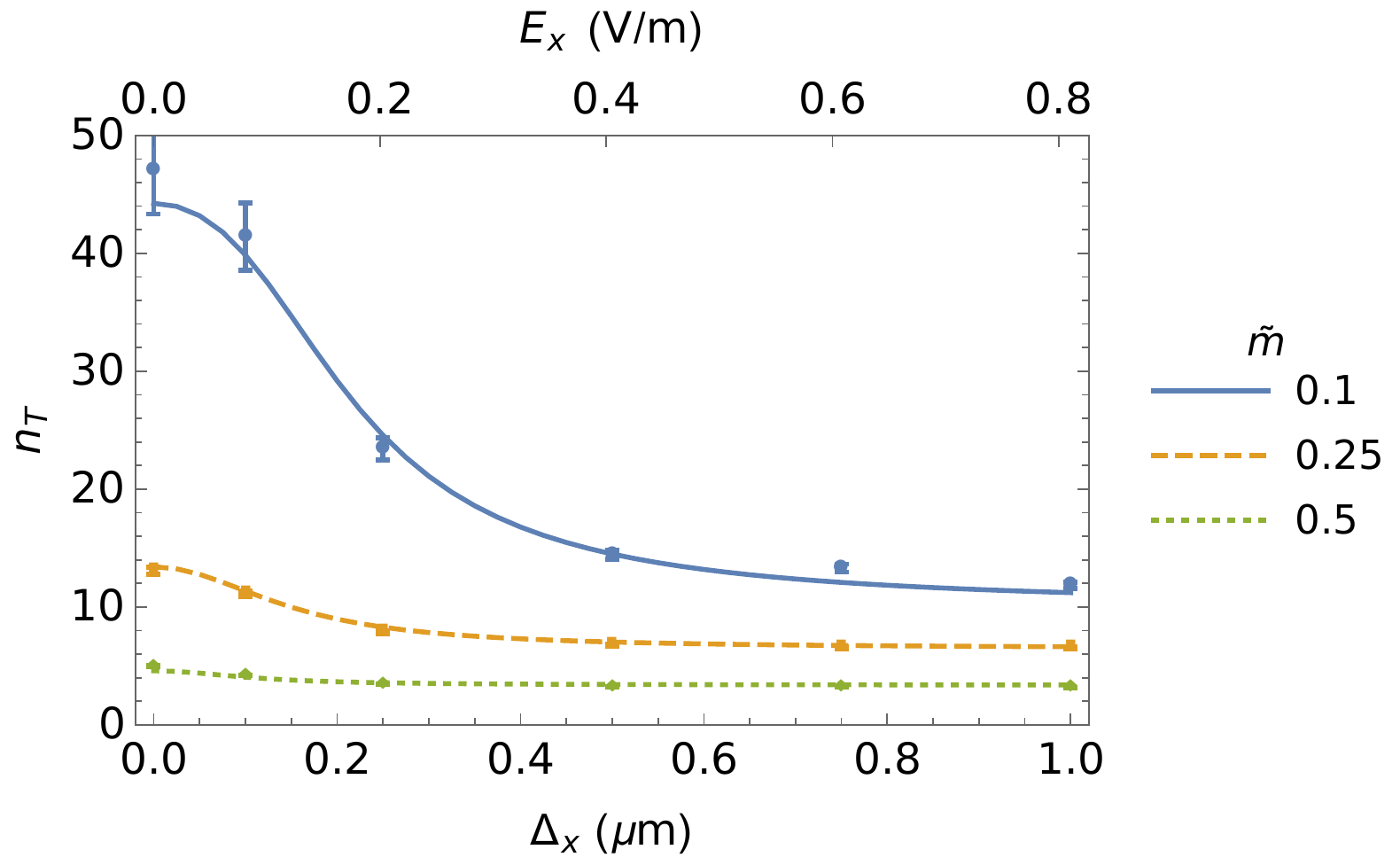}
\caption{(color online) The Tsallis exponent $n_T$ as a function of the applied electric field for a fixed buffer gas temperature of $50~\mu$K from numerical simulations (points) and the predicted trend calculated from the mean and mean-square energy, $\hat{n_T}$, (line) for $q_r = 0.1, a_z = 0.000625$ over a range of values of the neutral-to-ion mass ratio, $\tilde{m} = m_b/m_i$. Error bars show the estimated standard error. 200'000 iterations of the numerical simulation per data point. }
\label{fig:exponent_emmAndta}
\end{figure}

So far, we have considered only one of the two sources of additive fluctuations at a time. That is, either EMM is present and $T_b = 0$, or the buffer gas has a finite temperature and there is zero EMM. In this case, the exponent is independent of the magnitude of the fluctuations, since changing $T_b$ or $g_j$ while the other is set to zero is equivalent to multipling the energy by a constant which simply rescales the underlying distribution without changing its form, and so $n_T$ remains unchanged \cite{riley10a}. In the more realistic case in which both forced motion and a non-zero buffer gas temperature are present, the value of $n_T$ obtained depends on the relative proportions of each.  In Fig.~\ref{fig:exponent_emmAndta}, we show the results of applying an electric field of varying magnitude while keeping the   temperature of the buffer gas fixed at a non-zero value. 
It can be seen that the analytical predictions given by $\hat{n_T}$ are in good agreement with the numerical values obtained, and further that for a buffer gas at a temperature  $T_b = 50~\mu$K only a small electric field is required to tune the exponent from one limit to the other. As noted in Ref.~\cite{berkeland98a}, uniform electric fields of a magnitude $1$ V/m may easily develop during an ion trapping experiment and this is already sufficient to significantly alter the observed Tsallis exponent. Furthermore, since this effect applies even at very low mass ratios it cannot be assumed that in these cases the ion will exhibit a thermal distribution unless the EMM is compensated to a high degree of accuracy such that it contributes a negligible amount of energy compared to the thermal energy of the buffer gas.

Finally, let us briefly address the heating effect described in Ref.~\cite{cetina12a}, which arises due to the finite time of interaction between the ion and the atom during which the ion can be displaced in the rf field. By itself, this serves to produce a lower bound on the energy of the ion analogously to the effects of non-zero values of $T_b$ and $v_{f,j}(\tau)$. Moreover, it has been shown numerically and experimentally that, at a mass ratio of $\tilde{m} \approx 1$, this effect does not lead to a change in the observed power-law exponent \cite{meir16a}, in agreement with the results obtained here that at high mass ratio the power-law tail is a result of the multiplicative fluctuations. At low mass ratio, however, we have shown that $n_T$ is sensitive to the nature of the additive noise, and the heating effects due to long-range ion-atom interaction may alter the observed value of $n_T$ in this regime if it dominates over the other additive contributions.


\section{Summary and conclusions}
We have extended previous models of the ion-neutral collision process of an ion in a radiofrequency trap immersed in a buffer gas to take into account the motion of the ion due to external forces in addition to the trapping potential, providing analytical expressions for the mean steady-state energy of the ion, and confirmed that the distribution may be modelled by Tsallis statistics when this motion is present, in agreement with previous experimental findings \cite{meir16a}. We have demonstrated that at low neutral-to-ion mass ratio the effects of excess micromotion result in a lower value of the Tsallis exponent, i.e., a more pronounced power-law tail compared to the exponent observed with the same trapping parameters in the absence of forced motion. We have shown that this is a result of the non-thermal additive fluctuations due to the forced motion. Our results open the possibility for tuning the achieved energy distribution simply by applying a uniform electric field across the trapping region, allowing for deterministic control of a non-equilibrium steady state.  

\begin{acknowledgments}
We acknowledge funding from the University of Basel, the Swiss Nanoscience Institute project P1214 and the Swiss National Science Foundation as part of the National Centre of Competence in Research, Quantum Science \& Technology (NCCR-QSIT) and grant nr. 200020\_175533.
\end{acknowledgments}

\bibliography{forced_motion_new,Main-Oct17}
\appendix

\section{The inhomogenous damped Mathieu equation} \label{appendix:dampedMathieu}
The equation of motion for an ion in a rf trap in the presence of both an external force and damping is,
\begin{equation} \label{eq:mathieuDamped2}
\ddot r_j + 2 \mu_j \dot r_j + (a_j - 2 q_j \cos 2\tau) r_j = g_j(\tau) .
\end{equation}
Defining $r_j = e^{-\mu_j \tau} p_j $ and substituting this into Eq.~\eqref{eq:mathieuDamped2} results in an inhomogenous undamped Mathieu equation,
\begin{equation}\label{eq:mathieuPEquation}
\ddot p_j(\tau)  + (\tilde{a}_j  - 2 q_j \cos 2\tau) p_j(\tau) = g_j(\tau) e^{\mu_j \tau},
\end{equation} 
with $\tilde{a}_j  = a_j - \mu_j^2$. The general solution is given by,
\begin{equation} \label{eq:mathieuP}
p_j(\tau) = A_j [\mathrm{ce}(\tilde{a}_j ,q_j,\tau) \cos \phi_j - \mathrm{se}(\tilde{a}_j ,q_j,\tau) \sin \phi_j] + p_{d,j}(\tau),
\end{equation}
where $p_{f,j}(\tau)$ is found through the variation of parameters \cite{boyce17a},
\begin{align} \label{eq:variationMathieuP}
\begin{aligned}
 p_{f,j}(\tau) = - &\frac{ \mathrm{ce}(\tilde{a}_j  ,q_j,\tau)}{W_j} \int \mathrm{se}(\tilde{a}_j    ,q_j,\tau) e^{\mu_j  \tau} g_j(\tau) d\tau \\ + & \frac{  \mathrm{se}(\tilde{a}_j  ,q_j,\tau)}{W_j} \int \mathrm{ce}(\tilde{a}_j  ,q_j,\tau) e^{\mu_j  \tau} g_j(\tau) d\tau.
\end{aligned}
\end{align}
Therefore,
\begin{equation}
\begin{split}
r_j(\tau) =& A_j e^{-\mu_j \tau}[\mathrm{ce}(\tilde{a}_j  ,q_j,\tau) \cos \phi_j   -  \mathrm{se}(\tilde{a}_j  ,q_j,\tau) \sin \phi_j] \\ &+ r_{f,j}(\tau),
\end{split}
\end{equation}
where $r_{f,j}(\tau) = p_{f,j}(\tau) e^{-\mu_j \tau}$. As a first case, we take $g_j(\tau) = g_j$, i.e., a constant force leading to in-phase excess micromotion. Evaluating Eq.~\eqref{eq:variationMathieuP} using the Fourier series definitions of the Mathieu functions, then multiplying by $e^{-\mu_j \tau}$ to obtain $r_{f,j}(\tau)$ results in,
\begin{widetext}
\begin{equation}
r_{f,j}(\tau) =  \frac{g_j}{W_j} \sum_{m,n} \frac{c_{2m} c_{2n} }{ (\beta_j+2m)^2 + \mu_j ^2 } \big\{ (\beta_j+2m)\cos [ 2(m-n)\tau] - \mu_j  \sin [2(m-n)\tau]  \big\} ,
\end{equation}
\end{widetext}
from which it can be seen that the amplitude of motion does not decrease over time, although the damping does slightly alter the amplitude and introduces a term in quadrature phase with the rf drive. 

 Next, we take an oscillating external force $g_j(\tau) = g_j \sin (\alpha_j \tau + \varphi_j)$. In what follows we assume that this force is off-resonant, that is, $\alpha_{j} \neq \beta_j + 2m$ for any integer $m$, since in the resonant case the ion's trajectory is unstable. The resulting forced motion is,
\begin{widetext}
 \begin{equation}
\begin{split}
r_{f,j}(\tau) = \frac{g_j}{2 W_j} \sum_{m,n}       c_{2m}  c_{2 n} \bigg\{ \frac{ (\alpha_j +\beta_j +2 m) \sin [\tau  (\alpha_j +2 m-2 n) + \varphi_j]+\mu_j  \cos [\tau  (\alpha_j +2 m-2 n)  + \varphi_j]}{ \left(\alpha_j
   ^2+2 \alpha_j  (\beta_j +2 m)+\beta_j ^2+4 \beta_j  m+4 m^2+\mu_j ^2\right)} \\ +\frac{ (-\alpha_j +\beta_j +2 m) \sin [\tau  (\alpha_j -2 m+2
   n)  + \varphi_j]-\mu_j  \cos [\tau  (\alpha_j -2 m+2 n) + \varphi_j]}{  \left(\alpha_j ^2-2 \alpha_j  (\beta_j +2 m)+\beta_j ^2+4 \beta_j  m+4 m^2+\mu_j ^2\right)}  \bigg\} .
\end{split}
 \end{equation}
\end{widetext}
This, again, does not exhibit a decay over time. The largest term of this motion is typically for $m=0,n=0$, and in the undamped case ($\mu_j = 0$) this produces,
\begin{equation}
r_{f,j}(\tau) \approx \frac{\beta_j g_j c_0^2}{W_j (\beta_j^2 - \alpha_j^2)} \sin (\alpha \tau +  \varphi_j).
\end{equation}
Thus, applying a position-independent oscillating force to the ion produces oscillations at the same frequency and in phase with this external force. The special case $\alpha_j = 2, \varphi_j = 0$ corresponds to an external force of the form $\sin \Omega t$, which in Ref.~\cite{berkeland98a} is used as an approximate model for the effects of a phase difference between rf electrodes. 

\section{Total kinetic energy of an ion in an rf trap} \label{appendix:energyConversion}
In the main text, the ion's energy is characterised in terms of the secular energy, which represents the energy associated with the lowest-frequency mode of motion. The procedure used to calculate the effects of a collision, however, requires only that this energy be proportional to $A_j^2$, and so also applies to the time-averaged kinetic energy of the intrinsic motion used in Ref.~\cite{chen14a}. Furthermore, for the purposes of, e.g., calculating  reaction rates the total time-averaged kinetic energy, including contributions from the secular motion, instrinsic micromotion, and forced motion, may be required, as this represents the kinetic energy avaliable during collisions.  The velocity of the ion is,
\begin{equation}
v_j(\tau) = A_j [\mathrm{\dot{ce_j}}(\tau)  \cos \phi_j  -  \mathrm{\dot{se_j}}(\tau)\sin \phi_j ] + v_{f,j}(\tau),
\end{equation}
where dots indicates the derivative with respect to $\tau$. To simplify the notation, we define,
\begin{equation}
v_{h,j}(\tau) = A_j [\cos \phi_j \mathrm{\dot{ce_j}}(\tau) - \sin \phi_j \mathrm{\dot{se_j}}(\tau)],
\end{equation}
where the index $h$ indicates that this is the solution to the homogenous equation. The average kinetic energy is given by \cite{chen14a},
\begin{equation} \label{eq:energyTotalFromAverage} 
E_{j,K} =    \frac{1}{2}  m_i  \frac{\Omega^2}{4} \Theta[ v_j(\tau)^2],
\end{equation}
where the prefactor of $\Omega^2/4$ handles the conversion from the units of time used in the Mathieu equation to SI units, and the operator $\Theta[h(\tau)]$ is defined by,
\begin{equation}
\Theta[h(\tau)] = \lim_{L\rightarrow \infty} \frac{1}{2L} \int_{-L}^{L} h(\tau) d\tau .
\end{equation}
We may write Eq.~\eqref{eq:energyTotalFromAverage} as,
\begin{equation}
E_{j,K} =   \frac{1}{2} m_i  \frac{\Omega^2}{4} (I_1 +2  I_2 + I_3),
\end{equation}
where,
\begin{equation}
I_1 =    \Theta[ v_{h,j}(\tau)^2] ,
\end{equation}
\begin{equation}
I_2 =      \Theta[ v_{h,j}(\tau)     v_{f,j}(\tau)] ,
\end{equation}
and,
\begin{equation}
I_3 = \Theta[ v_{f,j}(\tau)^2].
\end{equation}
To evaluate $I_1$, we use the Fourier series definitions of the Mathieu functions to write, 
\begin{equation}
v_{h,j}( \tau) =-A_j \sum_m c_{2m,j} (\beta_j + 2m) \sin[ (\beta_j + 2m)\tau + \phi_j ].
\end{equation}
Using this expression, we may evaluate $I_1$ term-by-term to produce,
\begin{equation}
I_1 = A_j^2 \frac{1}{2} \sum_m c_{2m,j}^2 (\beta_j + 2m)^2 .
\end{equation} 
Note that $\frac{1}{2} m_i  \frac{\Omega^2}{4} I_1$ corresponds to the time-averaged kinetic energy of the intrinsic motion and is proportional to $A_j^2$ \cite{chen14a}. For the ion's trajectory to remain bounded, the forced motion cannot contain any frequency components which coincide with the frequencies of the intrinsic motion \cite{boyce17a}.  That is, when expressed as a Fourier series, it cannot contain terms with frequencies given by $\beta + 2m$ for any integer $m$. Hence, when $v_{f,j}$ is written in terms of a Fourier series and substituted into $I_2$, this integral must average to zero due to the orthogonality of sine and cosine functions \cite{olver2010a}.  The third integral cannot be evaluated without specifying the external force and so we shall simply denote this result as $\overline{v_{f,j}^2}$. Thus,
\begin{equation}
E_{j,K} =   \frac{1}{2} m_i  \sum_m \frac{\Omega^2}{4} [A_j^2 \frac{1}{2} c_{2m,j}^2 (\beta_j + 2m)^2]  + \overline{v_{f,j}^2} .
\end{equation}
 Recall that the secular energy of the ion is given by $E_j =  \frac{m_i}{2} \frac{\Omega^2}{4} A_j^2 \beta^2_j c_{0,j}^2$. Hence,
\begin{equation}\label{eq:totalEnergyFromSecular}
E_{j,K} = \frac{1}{2} E_j \sum_m \frac{ c_{2m,j}^2 (\beta_j + 2m)^2}{\beta_j^2 c_{0,j}^2} + \frac{m_i}{2}  \frac{\Omega^2}{4} \overline{v_{f,j}^2} .
\end{equation}
Since Eq.~\eqref{eq:totalEnergyFromSecular}  is a linear function of $E_j$, we may obtain the ensemble average simply by replacing $E_j$ by $\langle E_j \rangle$, which is obtained as described in the main text. 

\section{Numerical methods} \label{appendix:numerical}
The numerical simulations were implemented in a C++ program and were performed via matrix propagation for the reasons of speed and computational accuracy as described in Ref.~\cite{devoe09a}, adapated to take into account the motion due to an additional, spatially-independent force  \cite{boyce17a}. For the simulations performed in this work, the collision rate is a constant but this may be altered to model a varying collision rate due to an energy-dependent cross-section or a non-uniform buffer gas density distribution \cite{zipkes11a,hoeltkemeier16a}. The density and temperature of the buffer gas is fixed for these simulations, see \cite{zipkes11a} for a discussion of how they may be updated after each collision to model the heating of the buffer gas by the ion. The Mathieu functions were evaluated up to the $m=\pm 5$ Fourier terms with coefficients calculated using Miller's algorithm, and  the characteristic exponent was found through numerical integration \cite{fruchting69a, olver2010a}.  The energy drift in the absence of collisions after 300 propagations was found to be $E_{300}/E_0 < 10^{-5}$ for $q=0.5$. The extraction of $n_T$ from numerically calculated values of the energy was performed using maximum likelihood estimation to avoid the systematic errors introduced by performing linear regression on the tail of the binned data, and furthermore eliminating the need to choose appropriate bin sizes and a cutoff point \cite{shalizi07a,clauset09a}. This estimation treats $k,n_T,\langle \beta \rangle$ as free parameters to be found from the unbinned data and is performed using Mathematica \cite{mathematica10}. The errors on the parameters found via MLE are calculated from the estimated Fisher matrix \cite{riley10a}.   The analytical expressions for the mean energies were evaluated using Mathematica's built-in implementations of the Mathieu functions (see supplemental materials), which were also used to validate the implementations in the C++ program.

\end{document}